\RequirePackage{lineno}
\documentclass[prd,twocolumn,showpacs,amsmath,amssymb]{revtex4-1}
\usepackage{dcolumn}
\usepackage{bm}
\usepackage{rotating}
\usepackage{xcolor}
\usepackage{xspace}
\usepackage{graphicx}
\usepackage{epstopdf}

\usepackage[perpage,symbol*]{footmisc}
\def\@makefnmark{\hbox{\textsuperscript{\@thefnmark}}}

\renewcommand{\figurename}{Fig.}
\newcommand{\figref}[1]{\figurename~\ref{#1}}

\newcommand{\dEdx}{\ensuremath{\text{d}E/\text{d}x}\xspace}

\newcolumntype{x}[1]{>{\centering\arraybackslash\hspace{0pt}}p{#1}}

\begin{document}
\normalsize
\parskip=5pt plus 1pt minus 1pt
\title{Study of  $D^{+} \to K^{-} \pi^+ e^+ \nu_e$}

\author{
      M.~Ablikim$^{1}$, M.~N.~Achasov$^{9,f}$, X.~C.~Ai$^{1}$,
      O.~Albayrak$^{5}$, M.~Albrecht$^{4}$, D.~J.~Ambrose$^{44}$,
      A.~Amoroso$^{49A,49C}$, F.~F.~An$^{1}$, Q.~An$^{46,a}$,
      J.~Z.~Bai$^{1}$
% ...
      \\
      \vspace{0.2cm}
      (BESIII Collaboration)\\
      \vspace{0.2cm} {\it
        $^{1}$ Institute of High Energy Physics, Beijing 100049, People's Republic of China\\
% ...
$^{4}$ Bochum Ruhr-University, D-44780 Bochum, Germany\\
        $^{5}$ Carnegie Mellon University, Pittsburgh, Pennsylvania 15213, USA\\
% ...
$^{9}$ G.I. Budker Institute of Nuclear Physics SB RAS (BINP), Novosibirsk 630090, Russia\\
% ...
$^{44}$ University of Rochester, Rochester, New York 14627, USA\\
% ...
$^{46}$ University of Science and Technology of China, Hefei 230026, People's Republic of China\\
% ...
$^{49}$ (A)University of Turin, I-10125, Turin, Italy; (B)University of Eastern Piedmont, I-15121, Alessandria, Italy; (C)INFN, I-10125, Turin, Italy\\
% ...
$^{a}$ Also at State Key Laboratory of Particle Detection and Electronics, Beijing 100049, Hefei 230026, People's Republic of China\\
% ...
$^{f}$ Also at the Novosibirsk State University, Novosibirsk, 630090, Russia\\
% ...
      }
    \vspace{2cm}
}


\begin{abstract}
We present an analysis of the decay $D^{+} \to K^{-} \pi^+ e^+ \nu_e$
based on data collected by the BESIII experiment at the $\psi(3770)$ resonance.
Using a nearly background-free sample of 18262 events,
we measure the branching fraction
$\mathcal{B}(D^{+} \to K^{-} \pi^+ e^+ \nu_e) = (3.77 \pm 0.03 \pm 
0.08)\%$. For $0.8<m_{K\pi}<1.0$ GeV/$c^{2}$
the partial branching fraction is
$\mathcal{B}(D^{+} \to K^{-} \pi^+ e^+ \nu_e)_{[0.8,1.0]} = 
(3.39 \pm 0.03 \pm 0.08)\%$.
A partial wave analysis shows that the dominant $\bar K^{*}(892)^{0}$
 component is accompanied by an \emph{S}-wave contribution 
 accounting for $(6.05\pm0.22\pm0.18)\%$ of the total rate 
 and that other components are negligible.
The parameters of the $\bar K^{*}(892)^{0}$ resonance
and of the form factors based on the 
spectroscopic pole dominance predictions are also measured.
We also present a measurement of 
the $\bar K^{*}(892)^{0}$ helicity basis form factors in a model-independent way.
\end{abstract}

\let\newpage\relax\maketitle

\section{Introduction}

The semileptonic decay $D^{+} \to K^{-} \pi^+ e^+ \nu_e$, named $D_{e4}$ decay,
 has received particular attention due to the relative simplicity
of its theoretical description and the large branching fraction. The matrix 
element of $D_{e4}$ decay can be factorized as the product of
the leptonic and hadronic currents. This makes it a natural place to study 
the $K\pi$ system in the absence of interactions with
other hadrons, and to determine the hadronic transition form factors. In this paper 
the analysis is done mainly for two purposes:

i) Measure the different $K\pi$ resonant and non-resonant amplitudes that 
contribute to this decay, including \emph{S}-wave and radially
excited \emph{P}-wave and \emph{D}-wave components. Accurate measurements of these 
contributions can provide helpful information for 
amplitude analyses of \emph{D}-meson 
and \emph{B}-meson decays.

ii) Measure the $q^{2}$ dependent transition form factors in the $D_{e4}$ 
decay, where $q^{2}$ is the invariant mass squared of
the $e\nu_{e}$ system. This can be compared with  
hadronic model expectations and lattice QCD computations~\cite{Bernard:1991bz}.

The decay $D^{+} \to K^{-} \pi^+ e^+ \nu_e$ proceeds dominantly through the 
$\bar K^{*}(892)^{0}$ vector resonance. High statistics 
in this decay allow accurate measurements of the $\bar K^{*}(892)^{0}$
resonance parameters.
Besides this dominant process,
both FOCUS and BABAR have observed an \emph{S}-wave contribution with a 
fraction of about 6\% in this $D_{e4}$ decay~\cite{Link:2002ev,pwa2}.
In BABAR's parameterization, 
the $K\pi$ \emph{S}-wave with the isospin of $I=1/2$ was composed of a
non-resonant background term and the $\bar K_{0}^{*}(1430)^{0}$~\cite{pwa2}.
The \emph{S}-wave modulus was parameterized as a polynomial
dependence on the $K\pi$ mass for the non-resonant component 
and a Breit-Wigner shape for the $\bar K_{0}^{*}(1430)^{0}$.
The phase was parameterized based on measurements of the LASS 
scattering experiment~\cite{lass}.  It was described as a sum of the
background term $\delta_{\rm BG}^{1/2}$ and the $\bar
K_{0}^{*}(1430)^{0}$ term $\delta_{\bar K_{0}^{*}(1430)}^{0}$, 
where the mass dependence of $\delta_{\rm BG}^{1/2}$ was described by
means of an effective range parameterization.
 BABAR used it to fit the data over a $K\pi$
 invariant mass $m_{K\pi}$ range up to 1.6 GeV/$c^{2}$, showing that this 
parameterization
could describe the data well. In addition, they did a 
model-independent
measurement of the phase variation with $m_{K\pi}$, which agreed
well with the fit result based on the LASS parameterization.
In this paper we use BABAR's parameterization to describe the \emph{S}-wave, 
and performe a model-independent measurement of its phase  as well.

Another goal of this analysis is to describe
the $D^{+} \to K^{-} \pi^{+} e^{+} \nu_{e}$ decay in terms of
helicity basis form factors that
give the $q^{2}$ dependent amplitudes of the $K\pi$ system  in any
of its possible angular momentum states~\cite{jg korner}.
Traditionally, they are written as linear
combinations of vector and axial-vector form factors which are assumed
to depend on $q^{2}$ according to the 
spectroscopic pole dominance
(SPD) model~\cite{Link:2002wg, jg korner}. In this analysis we present two ways
to measure them. One way is to use the SPD model
to describe the form factors in the partial wave analysis
 (PWA) framework.
Another way is to perform a non-parametric measurement of the $q^{2}$
dependence of the helicity basis form factors using a weighting technique,
free from the SPD assumptions. This study will provide a
better understanding of the semileptonic decay dynamics.

\section{Experimental and analysis details}
\label{sec:select}
The analysis is based on the data sample of 2.93 $\rm fb^{-1}$~\cite{BESIII:2013iaa,BESIII:ISRPIPIPAPER} 
collected in $e^+e^-$ annihilations at the $\psi(3770)$ peak,
which has been accumulated with the BESIII detector operated at the double-ring 
 Beijing Electron-Positron Collider (BEPCII).
 
The BESIII detector~\cite{besiii} is designed 
approximately cylindrically symmetric around the interaction point,
covering 93\% of the solid angle.
Starting from its innermost component, the BESIII detector consists of a 43-layer 
Main Drift Chamber (MDC), a time-of-flight (TOF) system with 
two layers in the barrel region and one layer for each end-cap, 
and a 6240-cell CsI(Tl) crystal electromagnetic calorimeter
 (EMC) with both barrel and end-cap sections. The barrel components
 reside within a superconducting solenoidal magnet providing a 1.0~T magnetic 
 field aligned with the beam axis. Finally, a muon chamber (MUC)
 consisting of nine layers of resistive plate chambers is incorporated
 within the return yoke of the magnet. 
In this analysis, the MUC information is not used.
 The momentum resolution 
 for charged tracks in the MDC is 0.5\% for transverse 
 momenta of 1~GeV/c. The MDC also provides specific ionization
 (\dEdx) measurements for charged particles, with a resolution better than 6\%
 for electrons from Bhabha scattering. 
 The energy resolution for showers 
 in the EMC is 2.5\% for 1~GeV photons. The time resolution of the TOF
 is 80~ps in the barrel and 110~ps in the endcaps.

A GEANT4-based detector simulation~\cite{geant}
is used to study the detector performance.
The production of the $\psi(3770)$ resonance is 
simulated by the generator KKMC ~\cite{kkmc}, which takes the 
beam energy spread and the initial-state radiation (ISR) into account.
The decays of Monte-Carlo (MC) events are generated with EvtGen~\cite{evtgen}.
The final-state radiation (FSR) of charged particles is considered with 
the PHOTOS package~\cite{photos}.
Two types of MC samples are involved in this analysis: ``generic  MC" 
and ``signal MC". Generic  MC  consists of 
$D\bar{D}$ and non-$D\bar{D}$ decays of $\psi(3770)$,
ISR production of low-mass $\psi$ states, and QED and 
$q\bar{q}$ continuum processes.
The effective luminosities of the above MC samples correspond to 
5 to 10 times those of the experimental data. All the known decay 
modes are generated with the branching fractions taken from 
the Particle Data Group (PDG)~\cite{PDG}, 
while the remaining unknown processes are simulated with LundCharm~\cite{lundcharm}.
Signal MC is produced to simulate exclusive 
$\psi(3770)\to D^{+}D^{-}$ decays, where $D^{+}$ decays 
to the semileptonic signals uniformly (named ``PHSP signal MC'')
or with the decay intensity distribution determined by PWA (named ``PWA signal MC''),
while $D^{-}$ decays inclusively as in generic MC.


We use the technique of tagged $D$-meson decays
\cite{dtag method}.
At 3.773 GeV annihilation energy $D$ mesons are produced
in pairs. If a decay of one $D$ meson (``tagged decay'')
has been fully reconstructed in an event, 
then the existence of another $\bar{D}$ decay (``signal decay'')
in the same event is guaranteed. The tagged decays are reconstructed 
in the channels with larger branching fractions and lower background levels.
Six decay channels are considered:
$D^- \to K^+ \pi^- \pi^-$, 
$D^- \rightarrow K^+ \pi^- \pi^- \pi^0$, 
$D^- \rightarrow K_S^0 \pi^-$, 
$D^- \rightarrow K_S^0 \pi^- \pi^0$,
$D^- \rightarrow K_S^0 \pi^- \pi^- \pi^+$,
 and $D^- \rightarrow K^+ K^- \pi^-$.
The event selection consists of several stages:
selection and identification of particles (tracks and electromagnetic showers),
selection of the tagged decays, and selection of the signal decays 
$D^{+} \to K^{-} \pi^+ e^+ \nu_e$. Throughout this
paper, unless explicitly stated otherwise, the charge
conjugate is also implied when a decay mode of a
specific charge is stated.

Good tracks of charged particles are selected
by the requirement that the track origin is close
to the interaction point (within 10 cm along the beam axis
and within 1 cm in the perpendicular plane), and that 
the polar angle $\theta$ between the track and the beam direction
is within the good detector acceptance, $|\cos{\theta}|<0.93$.
The photons used for the neutral pion reconstruction 
are selected as electromagnetic showers with a minimum energy of 25 MeV 
in the barrel region 
($|\cos{\theta}|<0.8$) or 50 MeV in the endcaps 
($0.86<|\cos{\theta}|<0.92$). The shower timing measured 
by the calorimeter has to be within 700~ns after the beam collision.

Charged particle identification (PID) for pions
and kaons is based on the 
combined measurements of the d$E$/d$x$ and TOF.
Hypotheses for the track to be pion or kaon are considered.
Each track is characterized by $P(\pi)$ and $P(K)$,
which are the likelihoods for the pion and kaon hypotheses.
The pion candidates are identified with the requirement $P(\pi)>P(K)$
and the kaon candidates are required to have $P(K)>P(\pi)$.

The electron identification includes the measurements of
the energy deposition in the EMC in
addition to the \dEdx and TOF information. The measured values are
used to calculate the likelihoods	 $P_2$ for different particle 
hypotheses. The electron candidates
have to satisfy the following criteria:
 $P_2(e)/((P_2(K)+P_2(\pi)+P_2(e)) > 0.8$, ~ 
$P_2(e) > 0.001$. Additionally, the EMC energy of the 
electron candidate has to be more than 80\% of the track momentum
measured in the MDC.

Neutral pions are reconstructed from pairs of good photons
with an invariant mass in the range 
$115<M_{\gamma\gamma}<150$ MeV/$c^{2}$
and with a $\chi^2$ value for the 1-C 
mass constrained kinematic fit of $\pi^0 \to \gamma\gamma$ less than 200.
Candidates with both photons from the EMC endcap regions are rejected.

Neutral $K_{S}^{0}$ candidates are reconstructed with pairs of 
 oppositely charged tracks 
which are constrained to have a common vertex. 
The tracks from the $K_{S}^{0}$ decay are not required to satisfy
the good track selection or PID criteria. 
 Assuming the two tracks to be pions, we require
they have an invariant mass in the range 
$487 <M_{\pi^{+}\pi^{-}}<511$ MeV/$c^{2}$. 
The closest approach of the  track should be 
within 20 cm from the interaction point along the beam direction 
and the polar angle has to satisfy $|\cos{\theta}|<0.93$.

Appropriate combinations of the charged tracks and 
photons are formed for the six tagged $D^-$ decay channels.
Two variables
are calculated for each possible track combination: 
$M_{\rm BC} = \sqrt{E_{\rm beam}^2 -  |\vec p_{D}|^2}$, $\Delta E = E_D - E_{\rm beam}$,
where $E_D$ and $\vec p_{D}$ are the reconstructed energy and momentum
of the $D^-$ candidate, and $E_{\rm beam}$ is the beam energy.
$\Delta E$ is required 
to be consistent with zero within approximately twice 
the experimental resolution,
while $M_{\rm BC}$ should be
within the signal region $1.863 < M_{\rm BC} < 1.877$ GeV/$c^{2}$.
In each event we accept at most one candidate per tag mode per charge;
in the case of multiple candidates, the one with the smallest
$\Delta E$ is chosen.

\begin{figure}[htp]
  \begin{center}
  \includegraphics[width=0.48\linewidth]{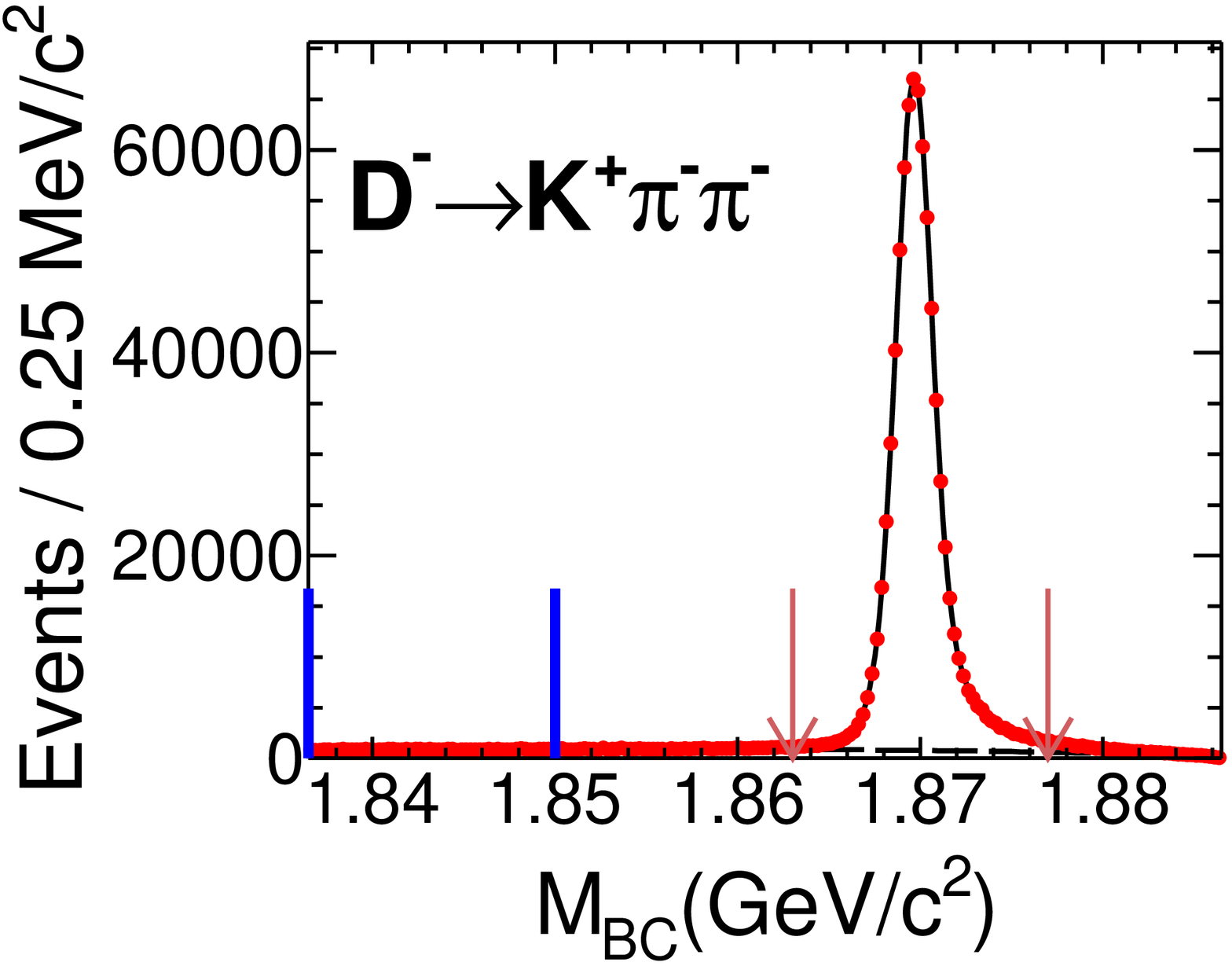}
  \includegraphics[width=0.48\linewidth]{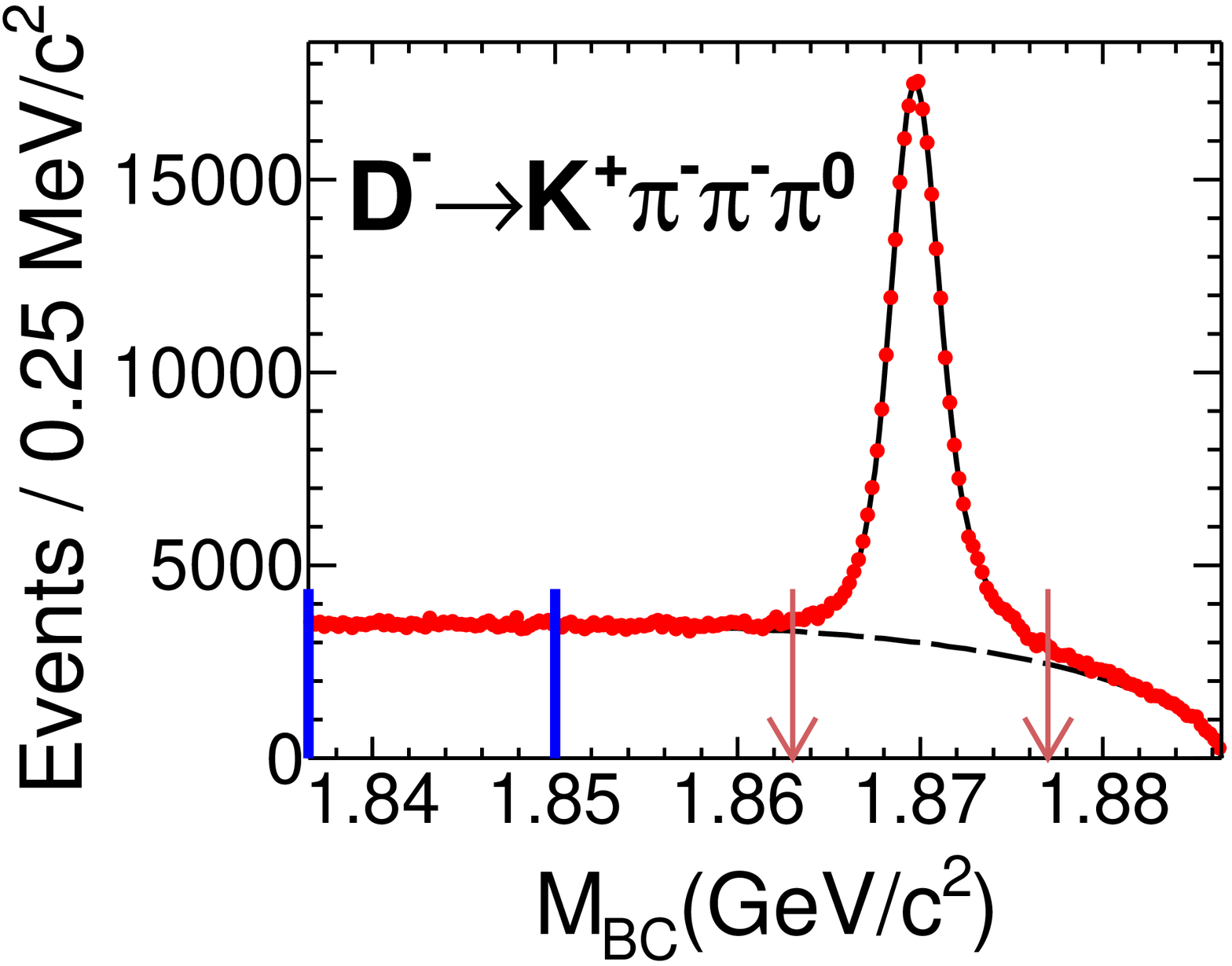} \\
  \includegraphics[width=0.48\linewidth]{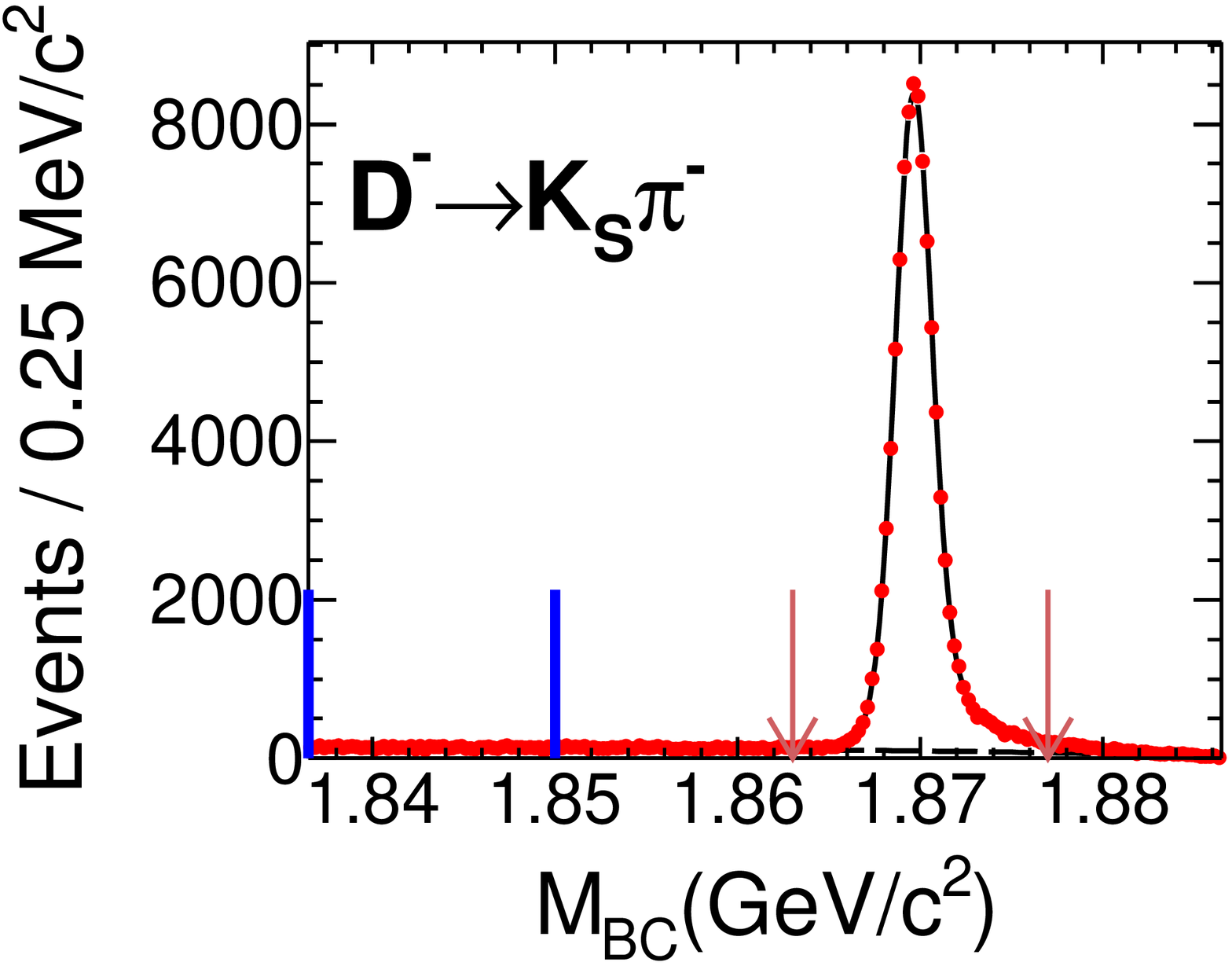}
  \includegraphics[width=0.48\linewidth]{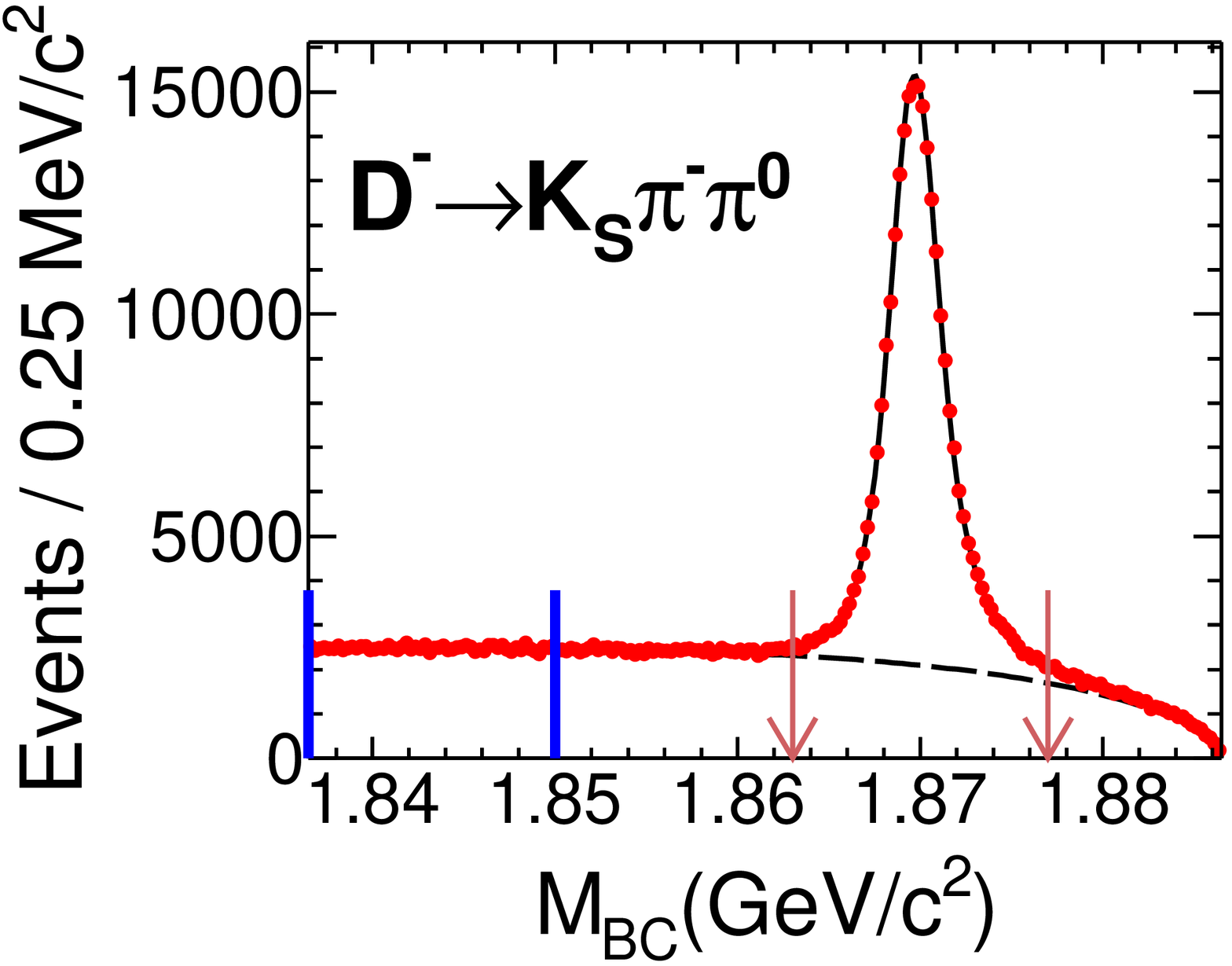} \\
  \includegraphics[width=0.48\linewidth]{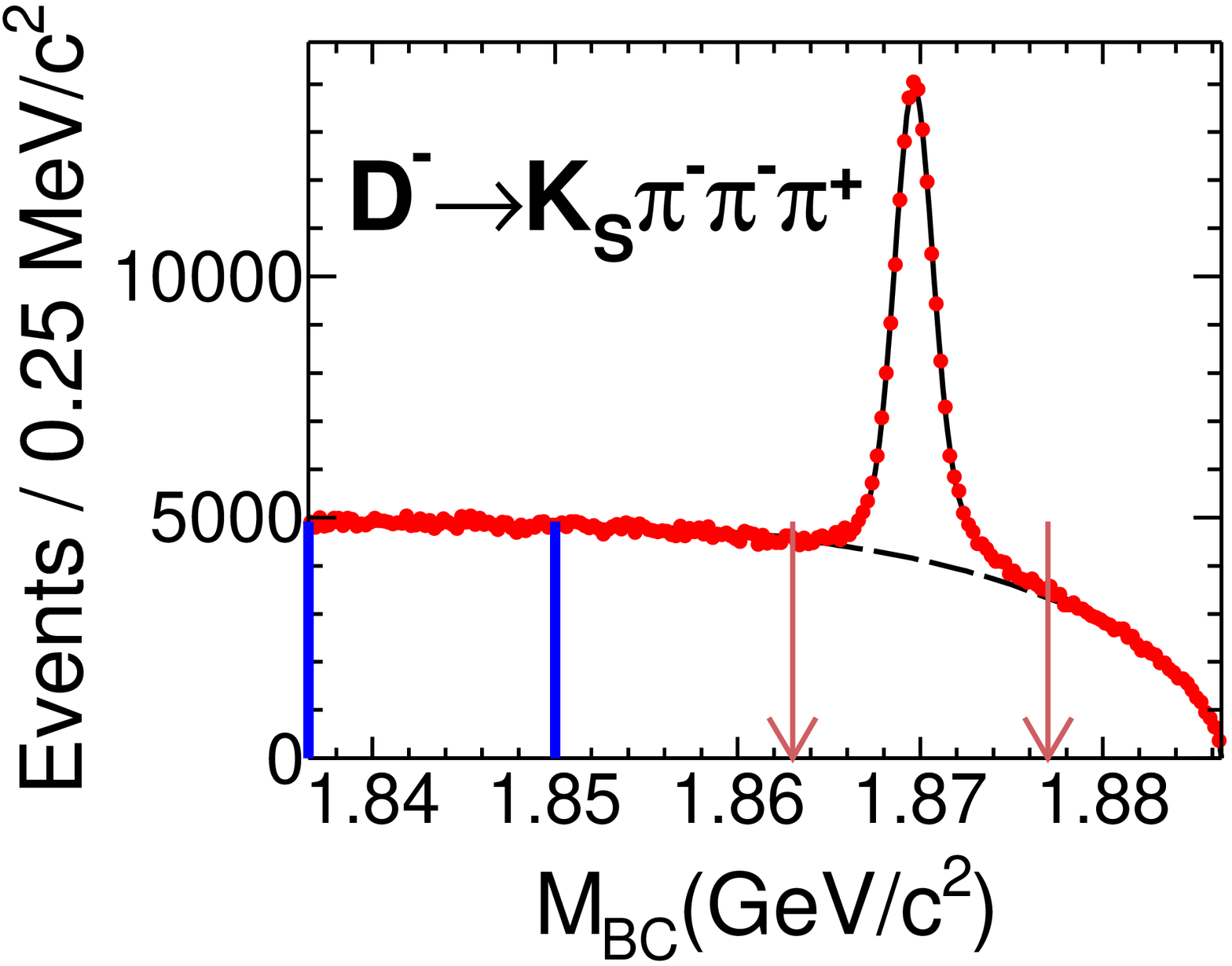}
  \includegraphics[width=0.48\linewidth]{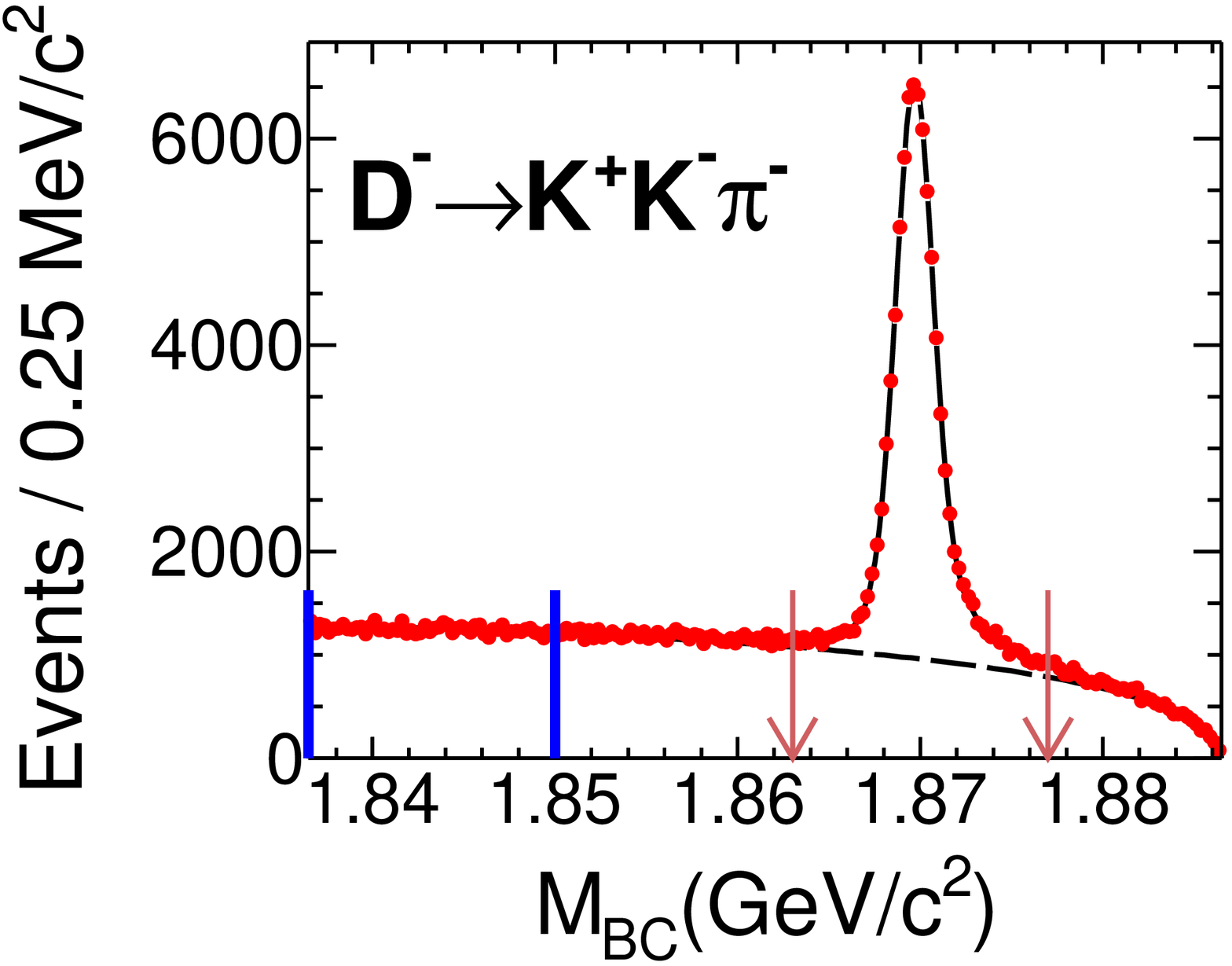}
  \caption{Fits to the $M_{\rm BC}$ distributions 
  for different tagged decay channels.
  The dots with error bars represent data and the 
  solid curves show the fits, which are the sum of signals and 
  background events. The background components are shown by the dashed lines.
  The areas between the arrows represent the signal regions
  while those between the vertical solid lines show the sidebands.}
  \label{fig:mbc data}
  \end{center}
\end{figure}

The tagged decay yields are determined separately
for the six tag channels. 
The yields are obtained by fitting the 
signal and background contributions to the $M_{\rm BC}$ distribution 
(\figref{fig:mbc data}) of 
the events passing the $\Delta E$ cuts. 
The signal shape is modeled by
the reconstructed MC distribution, while the background shape
is described by the ARGUS function~\cite{ARGUS}.
The yields are determined by 
subtracting the numbers of background events 
from the total numbers of events
in the $M_{\rm BC}$ signal region. 
The yields of the six tags $N_{\rm tag}$, 
together with the tag efficiencies $\epsilon_{\rm tag}$ 
estimated by generic MC,
 are listed in Table~\ref{tab:tags}.

The signal decay $D^{+} \to K^{-} \pi^+ e^+ \nu_e$
is reconstructed from the tracks remaining after the selection
of the $D^-$ tag. We require that there are 
exactly three tracks on the 
signal side satisfying the good track selection criteria,
and they must be identified 
as $K^{-}$, $\pi^{+}$ and $e^{+}$.

The energy $E_{\rm miss}$ and momentum $\vec{p}_{\rm miss}$
of the missing neutrino are reconstructed using energy 
and momentum conservation. Background events with an undetected 
massive particle are suppressed by the requirement
$|U_{\rm miss}|<0.04$ GeV, where $U_{\rm miss} = E_{\rm miss} - |\vec p_{\rm miss}|$.
The background from neutrino-less decays is suppressed
by the selection criterion $E_{\rm miss}>0.04$ GeV.

The background from the events containing neutral pions 
is suppressed by the requirement that no unassociated EMC 
shower has an energy deposition above 0.25 GeV.
Only the clusters separated by more than 15$^\circ$
from the closest charged tracks are considered.

Finally, in order to reject  cross-feed from the $e^+ e^- \to D^0 \bar{D}^{0}$ events,
an additional selection is applied to the events where
the tagged decay  is reconstructed in the channels 
$D^- \rightarrow K_S^0 \pi^- \pi^- \pi^+$,
$D^- \rightarrow K_S^0 \pi^- \pi^0$ and
$D^- \rightarrow K^+ \pi^- \pi^- \pi^0$. 
For such events reconstruction of a purely hadronic decay
of a neutral $D^0$ or $\bar{D}^0$ meson is attempted  using the tracks from the entire event.
The event is rejected if any $D^0$  candidate satisfies the tight selection 
criteria $1.860<M_{\rm BC}<1.875$ GeV/$c^2$ and $|\Delta E|<0.01$ GeV.

\begin{figure}[htp]
  \begin{center}
  \includegraphics[width=0.7\linewidth]{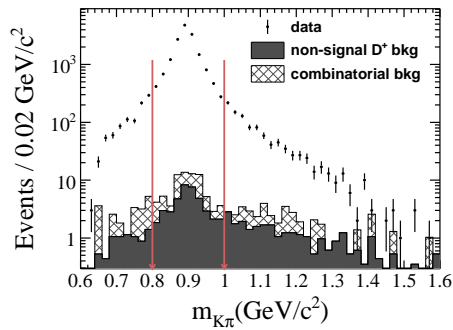}
  \caption{$m_{K\pi}$ distribution of the selected candidates. 
  The range between the arrows corresponds to the 
   $K^*$-dominated region.
  The dots with error bars represent data, the shadowed histogram
  shows the non-signal $D^+$ background estimated from MC simulation 
  and the hatched area shows the combinatorial background
  estimated from the $M_{\rm BC}$ sideband of data.}
  \label{fig:mkpi}
  \end{center}
\end{figure}

In total, 18262 candidates are selected (denoted as $N_{\rm obs}$).
The $m_{K\pi}$ distribution of these candidates is 
illustrated in \figref{fig:mkpi}
in the full $m_{K\pi}$ range 0.6$<$$m_{K\pi}$$<$1.6 GeV/$c^{2}$. 
In the $K^*$-dominated 
region  0.8$<$$m_{K\pi}$$<$1.0 GeV/$c^{2}$ (corresponding to 
the area between the arrows), 16181 candidates are located.

%
%
%
 
MC simulation shows that the background level is about 0.8\% 
over the full $m_{K\pi}$ range and around 0.5\% in the 
$K^*$-dominated region. 
The backgrounds can be divided into two categories.
One category arises from non-signal $D^+$ decays, including  
 $D^{+} \to K^- \pi^+ \pi^+ \pi^{0}$,
$D^{+} \to K^- \pi^+ \pi^+$ 
and $D^{+} \to K^- \pi^+ \mu^+ \nu_\mu$,
among which the last one is the largest
contribution, arising when $\mu^{+}$ is misidentified as $e^{+}$.
For the non-signal $D^+$ background, the accompanying 
$D^-$ meson peaks in the $M_{\rm BC}$ distribution
in the same way as  when $D^+$ decays to signals.
The number of this background 
 is estimated using MC simulation,
  76$\pm$3 over the full $m_{K\pi}$ range and 
40$\pm$2 in the $K^*$-dominated region
(The errors are statistical only).
The other category is combinatorial background, mainly 
due to $e^+e^- \to D^0\bar{D}^{0}$ events 
and the $e^+e^- \to q\bar{q}$ continuum.
This background has a continuum $M_{\rm BC}$ spectrum 
and can be estimated from data using the events
located in the sideband (see \figref{fig:mbc data}).
The scaled contribution from this background  is 69$\pm$7 and
 33$\pm$5 over the full $m_{K\pi}$ range and in 
the $K^*$-dominated region, respectively. 
The backgrounds from both categories are illustrated in 
 \figref{fig:mkpi} , and the total number
(denoted as $N_{\rm bkg}$) 
can be obtained by summing them up.

\section{Determination of the branching fraction}
\label{sec:BF}

The branching fraction of the decay $D^{+} \to K^{-} \pi^+ e^+ \nu_e$ 
is calculated using

\noindent
\begin{equation}
\mathcal{B}_{\rm sig}=\frac{N_{\rm obs}-N_{\rm bkg}}
{\sum_{\alpha}N_{\rm tag}^{\alpha}
\epsilon_{\rm tag,sig}^{\alpha}/\epsilon_{\rm tag}^{\alpha}},
\label{eq:br}
\end{equation}

\noindent
where $N_{\rm obs}$ and $N_{\rm bkg}$ are the numbers of the 
observed  and the background events (see Sec.~\ref{sec:select}).
For the tag mode $\alpha$, 
$N_{\rm tag}^{\alpha}$  is the number of the tagged $D^-$ mesons,
$\epsilon_{\rm tag}^{\alpha}$ is the reconstruction efficiency,
and $\epsilon_{\rm tag,sig}^{\alpha}$ represents the combined  efficiency to reconstruct 
both $D^+$ and $D^-$.

The selection efficiency $\epsilon_{\rm tag,sig}$  depends significantly
on the relative contribution of different $(K\pi)$ states.
Therefore, we exploit two ways to calculate the branching fraction.
One way is to use the PWA method to 
estimate precisely the contributions from different processes
in the $D^{+} \to K^{-} \pi^+ e^+ \nu_e$ 
final state. $\epsilon_{\rm tag,sig}$ is determined
by signal MC
 which is based on the PWA results.
Another way is to determine the branching fraction
in the $K^*$-dominated region.
This region is dominated by the $\bar{K}^{*}(892)^{0}$ resonance
and the determination of the branching fraction 
is nearly independent of the model describing the composition 
of the decay.

The PWA procedure will be described in detail in Sec.~\ref{sec:pwa}.
The selection efficiencies $\epsilon_{\rm tag,sig}$ for both the methods 
are summarized in Table~\ref{tab:tags}. 
 The resulting branching fractions are obtained 
over the full $m_{K\pi}$ range
and in the $K^*$-dominated region as
 \begin{eqnarray}
\mathcal{B}(D^{+} \to K^{-} \pi^+ e^+ \nu_e)=(3.77 \pm 0.03 \pm 0.08)\%,
\label{eq:br_total}
\\
\mathcal{B}(D^{+} \to K^{-} \pi^+ e^+ \nu_e)_{[0.8,1.0]}=(3.39 \pm 0.03 \pm 0.08)\%,
\label{eq:br_kstar}
\end{eqnarray}
where the first errors are statistical and the second are systematic.

\begin{table*}[htp]
  	\renewcommand{\arraystretch}{1.1}
  \renewcommand{\tabcolsep}{0.01\linewidth}
  \begin{center}
  \caption{Summary of event selection for different tag modes,
  where the errors are statistical.}
  \begin{tabular}{c|c|c|c|c}
    \hline
    \hline
    Tag  &  $N_{\rm tag}$  &
             $\epsilon_{\rm tag}$ (\%) &
             $\epsilon_{\rm tag,sig}$ (\%) & 
             $\epsilon_{\rm tag,sig}$ (\%)\\
&  & &full $m_{K\pi}$ range  & $K^*$-dominated region 
\\ 
    \hline   
    $K^{+}\pi^{-}\pi^{-}$      & 776648$\pm$915 &  50.62$\pm$0.02  & 
    16.46$\pm$0.02  &      16.30$\pm$0.02  \\

    $K^{+}\pi^{-}\pi^{-}\pi^{0}$ & 234979$\pm$678 & 25.23$\pm$0.02 &  
     7.71$\pm$0.02  &  7.62$\pm$0.02   \\

    $K_{S}^{0}\pi^{-}$      & 95498$\pm$320 & 53.91$\pm$0.06 &  
    17.55$\pm$0.07   & 17.34$\pm$0.07  \\

    $K_{S}^{0}\pi^{-}\pi^{0}$ & 215619$\pm$610  & 29.24$\pm$0.03 &   
      9.06$\pm$0.02 &  8.95$\pm$0.02   \\

    $K_{S}^{0}\pi^{-}\pi^{-}\pi^{+}$ & 120491$\pm$648  & 37.33$\pm$0.06 & 
     11.55$\pm$0.04   & 11.00$\pm$0.04  \\

    $K^{-}K^{+}\pi^{-}$          & 69909$\pm$374  & 40.78$\pm$0.07 & 
     13.18$\pm$0.06   &  13.04$\pm$0.06   \\
    \hline  
    \hline   
  \end{tabular}
    \label{tab:tags}  
  \end{center}
\end{table*}

The largest contributions to the systematic uncertainties for the branching fraction
originate from the MC determination of the efficiencies
of track reconstruction (1.73\%) and particle identification (0.95\%).
They are estimated using clean samples of pions, kaons and electrons. 

The uncertainties due to the selection criteria are estimated by comparing the corresponding selection efficiencies 
between data and MC using clean control samples. 
The uncertainty due to the $U_{\rm miss}$ requirement (0.76\%) is estimated using fully-reconstructed $D^+ \to K^{-} \pi^{+} \pi^{+}$, 
$D^- \to K^{+} \pi^{-} \pi^{-} \pi^0$ decays by treating one photon as a missing particle.
The uncertainty due to the selection on the electron $E/p$ ratio
(0.36\%) is obtained using electrons from radiative Bhabha scattering.
To obtain the uncertainty due to the shower isolation requirement (0.26\%),
fully reconstructed $D^{+} \to K^{-} \pi^{+} \pi^{+}$, $D^{-} \to K^{+} \pi^{-} \pi^{-}$ decays are used.

We vary the $M_{\rm BC}$ fit range to  estimate the associated uncertainty (0.32\%).
We also consider uncertainties due to
imperfections of the PWA model (0.23\%). This is
estimated by varying parameters in the 
probability density function (PDF, the detail of which will be described  in Eq.~\eqref{eq:l})
by $1\,\sigma$ 
and considering additional resonances. 
To estimate the uncertainty due to the background  fraction (0.16\%),
we change the branching fractions by $1\,\sigma$
according to PDG for the non-signal $D^+$ background and vary the normalization
by $1\,\sigma$ for the combinatorial background. 
As for the uncertainty due to the shape of the background distribution (0.12\%), 
only the uncertainty from the $D^{+}\to K^{-}\pi^{+}\pi^{+}\pi^{0}$  background
is non-negligible, which is estimated by comparing
the difference between two extreme cases: phase space process and 
$D^{+}\to \bar{K}^{*}(892)^{0}\rho^{+}$.

The total systematic uncertainties are calculated by adding 
the above uncertainties in quadrature, resulting in 2.21\% for both the branching fraction 
over the full $m_{K\pi}$ range and in the $K^*$-dominated region.

\section{PWA of $\bm{D^+ \to K^- \pi^+ e^+ \nu_e}$ decay}
\label{sec:pwa}

\begin{figure}[htp]
  \begin{center}
  \includegraphics[width=0.6\linewidth]{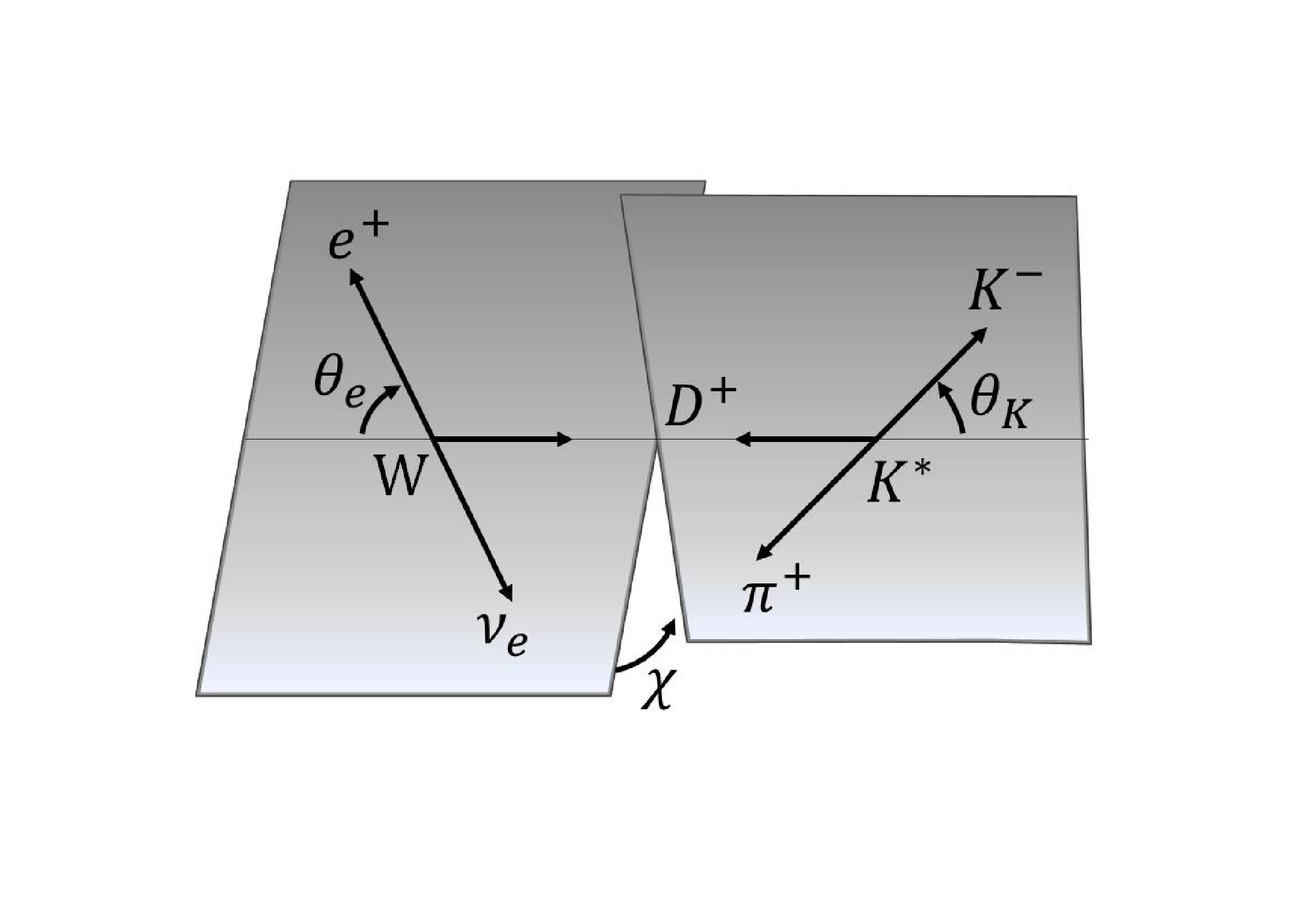}
  \caption{Definition of the angular variables.}
  \label{fig:phys_angle}
  \end{center}
\end{figure}

The four-body decay $D^+ \to K^- \pi^+ e^+ \nu_e$  can be uniquely described 
by the five kinematic variables~\cite{pwa5}: 
$K\pi$ mass square ($m^{2}$), $e\nu_{e}$ mass square ($q^{2}$), the angle 
between the $\pi$ and the $D$ direction in the $K\pi$ rest frame ($\theta_{K}$), 
the angle between the $\nu_{e}$ and the $D$ 
direction in the $e\nu_{e}$ rest frame ($\theta_{e}$), 
and the angle between the two decay planes ($\chi$).
The angular variables are illustrated in  \figref{fig:phys_angle}.
The sign of $\chi$ 
should be changed when analyzing $D^{-}$ in order to maintain $CP$ conservation.

Neglecting the mass of $e^+$,
the differential decay width can be expressed as:

\noindent
\begin{equation}
  \begin{aligned}
  d^{5}\Gamma =& \frac{G^{2}_{F}|V_{cs}|^{2}}{(4\pi)^{6}m^{3}_{D}}X\beta \mathcal{I}(m^{2}, q^{2}, \theta_{K}, \theta_{e}, \chi) \\
  & \times  dm^{2} dq^{2} d\cos(\theta_{K}) d\cos(\theta_{e})d\chi, \\
  X=& p_{K\pi}m_{D}, \quad  \beta=2p^{*}/m,
  \label{eq:decay_rate}
  \end{aligned}
\end{equation}

\noindent
where $G_{F}$ is the Fermi constant, 
 $V_{cs}$ is the $c$$\to$$s$ element of the 
Cabibbo-Kobayashi-Maskawa matrix,
 $p_{K\pi}$ is the momentum of the $K\pi$ system in the $D$ rest frame, 
 and $p^{*}$ is the momentum of the $K$ in the $K\pi$ 
rest frame. The dependence of the decay intensity $\mathcal{I}$ on $\theta_{e}$ 
and $\chi$ is given by Ref.~\cite{pwa3}:

\noindent
\begin{equation}
  \begin{aligned}
\mathcal{I}=& \mathcal{I}_{1}+\mathcal{I}_{2}\cos2\theta_{e}+\mathcal{I}_{3}\sin^{2}\theta_{e}\cos2\chi+\mathcal{I}_{4}
  \sin2\theta_{e}\cos\chi  \\
  &+\mathcal{I}_{5}\sin\theta_{e}\cos\chi +\mathcal{I}_{6}\cos\theta_{e}+\mathcal{I}_{7}\sin\theta_{e}\sin\chi 
  \\
  &+\mathcal{I}_{8}\sin2\theta_{e}\sin\chi
  +\mathcal{I}_{9}\sin^{2}\theta_{e}\sin2\chi,
  \label{eq:decay_intensity}
  \end{aligned}
\end{equation}  

\noindent
where $\mathcal{I}_{1, \ldots,9}$ depend on $m^{2}$, $q^{2}$, and $\theta_{K}$. These 
quantities can be expressed in terms of the three form factors $\mathcal{F}_{1,2,3}$:

\noindent
\begin{equation}
  \begin{aligned}
  \mathcal{I}_{1}=&\frac{1}{4}\{|\mathcal{F}_{1}|^{2}+\frac{3}{2}\sin^{2}\theta_{K}
  (|\mathcal{F}_{2}|^{2}+|\mathcal{F}_{3}|^{2})\}, \\
  \mathcal{I}_{2}=&-\frac{1}{4}\{|\mathcal{F}_{1}|^{2}-\frac{1}{2}\sin^{2}\theta_{K}
  (|\mathcal{F}_{2}|^{2}+|\mathcal{F}_{3}|^{2})\}, \\
  \mathcal{I}_{3}=&-\frac{1}{4}\{|\mathcal{F}_{2}|^{2}-|\mathcal{F}_{3}|^{2}\}\sin^{2}\theta_{K}, \\
  \mathcal{I}_{4}=&\frac{1}{2}\rm{Re}(\mathcal{F}^{*}_{1}\mathcal{F}_{2})\sin\theta_{K}, \\
  \mathcal{I}_{5}=&\rm{Re}(\mathcal{F}^{*}_{1}\mathcal{F}_{3})\sin\theta_{K}, \\
  \mathcal{I}_{6}=&\rm{Re}(\mathcal{F}^{*}_{2}\mathcal{F}_{3})\sin^{2}\theta_{K}, \\
  \mathcal{I}_{7}=&\rm{Im}(\mathcal{F}_{1}\mathcal{F}_{2}^{*})\sin\theta_{K}, \\
  \mathcal{I}_{8}=&\frac{1}{2}\rm{Im}(\mathcal{F}_{1}\mathcal{F}_{3}^{*})\sin\theta_{K}, \\
  \mathcal{I}_{9}=&-\frac{1}{2}\rm{Im}(\mathcal{F}_{2}\mathcal{F}^{*}_{3})\sin^{2}\theta_{K}.
  \end{aligned}
\end{equation}

Then one can expand $\mathcal{F}_{i=1,2,3}$ into partial waves including
 \emph S-wave ($\mathcal{F}_{10}$), \emph P-wave ($\mathcal{F}_{i1}$)
and \emph D-wave ($\mathcal{F}_{i2}$):

\noindent
\begin{equation}
  \begin{aligned}
  \mathcal{F}_{1}&=\mathcal{F}_{10}+\mathcal{F}_{11}\cos\theta_{K}+\mathcal{F}_{12}\frac{3\cos^{2}\theta_{K}-1}{2}, \\
  \mathcal{F}_{2}&=\frac{1}{\sqrt{2}}\mathcal{F}_{21}+\sqrt{\frac{3}{2}}\mathcal{F}_{22}\cos\theta_{K}, \\
  \mathcal{F}_{3}&=\frac{1}{\sqrt{2}}\mathcal{F}_{31}+\sqrt{\frac{3}{2}}\mathcal{F}_{32}\cos\theta_{K}.
  \label{eq:form_factor}
  \end{aligned}
\end{equation}

\noindent
Here the parameterizations of $\mathcal{F}_{ij}$ are taken from the
BABAR collaboration~\cite{pwa2}.
Contributions with higher angular momenta are neglected. 

The \emph P-wave related form factors $\mathcal{F}_{i1}$ are parameterized by 
the helicity basis form factors $H_{0,\pm}$: 

\noindent
\begin{equation}
  \begin{aligned}
  \mathcal{F}_{11}=&2\sqrt{2}\alpha qH_{0}\times \mathcal{A}(m),  \\
  \mathcal{F}_{21}=&2\alpha q(H_{+}+H_{-})\times \mathcal{A}(m), \\
  \mathcal{F}_{31}=&2\alpha q(H_{+}-H_{-})\times \mathcal{A}(m). \\
  \label{eq:form_factor_P}
  \end{aligned}
\end{equation} 

\noindent
Here  $\mathcal{A}(m)$ denotes the amplitude characterizing the shape
of the resonances, which has a Breit-Wigner form  defined in Eq~(\ref{eq:BW}).
$\alpha$ is a constant factor given in Eq.~\eqref{eq:alpha},
which depends on the definition of $\mathcal{A}(m)$.
 The factorization in Eq.~\eqref{eq:form_factor_P} and in the following 
Eq.~\eqref{eq:form_factor_S} and Eq.~\eqref{eq:form_factor_D}
is based on the assumption that the $q^{2}$ dependence of the resonance
amplitude is weak for the narrow Breit-Wigner structure.
The helicity basis form factors can be related to 
one vector $V(q^{2})$ and two axial-vector $A_{1,2}(q^{2})$
form factors:

\noindent
\begin{equation}
  \begin{aligned}
  H_{0}(q^{2},m^{2})=&\frac{1}{2mq}[(m_{D}^{2}-m^{2}-q^{2})(m_{D}+m)A_{1}(q^{2}) \\
  &-4\frac{m_{D}^{2}p_{K\pi}^{2}}{m_{D}+m}A_{2}(q^{2})], \\
  H_{\pm}(q^{2},m^{2})&=[(m_{D}+m)A_{1}(q^{2})\mp\frac{2m_{D}p_{K\pi}}{(m_{D}+m)}V(q^{2})].
  \label{eq:helicity}
  \end{aligned}
\end{equation}

The $q^{2}$  dependence is expected to be determined by the 
singularities nearest to the $q^{2}$ physical region [0,$q^{2}_{max}$]
($q^{2}_{max}$$\sim$1.25 GeV$^{2}$/c$^{4}$),
which are assumed to be poles 
corresponding to the lowest vector ($D^{*}_{S}$) and axial-vector  ($D_{S1}$) states
for the vector and axial-vector  form factor, respectively.
We use the SPD model to describe the $q^{2}$ dependence:

\noindent
\begin{equation}
  \begin{aligned}
      V(q^{2})&=\frac{V(0)}{1-q^{2}/m_{V}^{2}},  \\
  A_{1}(q^{2})&=\frac{A_{1}(0)}{1-q^{2}/m_{A}^{2}},  \\
  A_{2}(q^{2})&=\frac{A_{2}(0)}{1-q^{2}/m_{A}^{2}},  
  \label{eq:spd}
  \end{aligned}
\end{equation}

\noindent
where $m_{V}$ and $m_{A}$ are expected to be close 
to $m_{D_{S}^{*}}\backsimeq$ 2.1 GeV/$c^{2}$ 
and $m_{D_{S1}}\backsimeq$ 2.5 GeV/$c^{2}$, respectively. 
In this analysis, the values of $m_{V}$, $m_{A}$ 
and the ratios of the form factors 
taken at $q^{2}=0$, $r_{V}=V(0)/A_{1}(0)$ 
and $r_{2}=A_{2}(0)/A_{1}(0)$, are determined by the PWA fit.
The value of $A_{1}(0)$ is determined by measuring the 
branching fraction of $D^{+}\to \bar{K}^{*}(892)^{0}e^{+}\nu_{e}$.

For the amplitude of the resonance $\mathcal{A}(m)$, we use a Breit-Wigner shape
with a mass-dependent width:

\noindent
\begin{equation}
  \mathcal{A}(m)=\frac{m_{0}\Gamma_{0}F_{J}(m)}
  {m_{0}^{2}-m^{2}-im_{0}\Gamma(m)},
  \label{eq:BW}
\end{equation}

\noindent
where $m_{0}$ and $\Gamma_{0}$ are the pole 
mass and total width of the resonance, respectively.
This parameterization is applicable to resonances 
of different angular momenta denoted by $J$.
In the case of the  \emph P-wave, $J=1$.
The mass-dependent width $\Gamma(m)$ is given by
\begin{eqnarray}
  \Gamma(m)=\Gamma_{0}\frac{p^{*}}{p^{*}_{0}}
  \frac{m_{0}}{m}F_{J}^{2}(m), \\
  \label{eq:width}
   F_{J}=\left(\frac{p^{*}}{p^{*}_{0}}\right)^{J}\frac{B_{J}(p^{*})}{B_{J}(p^{*}_{0})}.
  \label{eq:width_F}
\end{eqnarray}  

\noindent
Here $p^{*}$ is the momentum of the $K$ in the $K\pi$ rest frame, and 
$p^{*}_{0}$ is its value 
determined at $m_{0}$, the pole mass of the resonance.
$B_{J}$ is the Blatt-Weisskopf damping factor given by the following
expressions:

\noindent
\begin{equation}
  \begin{aligned} 
  B_{0}(p)&=1, \\
  B_{1}(p)&=1/\sqrt{1+r_{BW}^{2}p^{2}},  \\
  B_{2}(p)&=1/\sqrt{(r_{BW}^{2}p^{2}-3)^{2}+9r_{BW}^{2}p^{2}}.
  \label{eq:b}
  \end{aligned}
\end{equation}  

\noindent
The barrier factor $r_{BW}$, as well as $m_{0}$ and $\Gamma_{0}$ for $\bar{K}^{*}(892)^{0}$,
are free parameters in the PWA fit.

With the definition of the mass distribution 
given in Eq.~(\ref{eq:BW}), the factor $\alpha$ 
entering  Eq.~(\ref{eq:form_factor_P}) is given by

\noindent
\begin{equation}
  \alpha = \sqrt{\frac{3\pi \mathcal{B}_{K^*}}{p^{*}_{0}\Gamma_{0}}},
  \label{eq:alpha}
\end{equation}

\noindent
where $\mathcal{B}_{K^*}=\mathcal{B}({K^{*}\to K^-\pi^+})=2/3$.

The \emph S-wave related form factor $\mathcal{F}_{10}$ is expressed as

\noindent
\begin{equation}
  \mathcal{F}_{10}=p_{K\pi}m_{D}\frac{1}{1-\frac{q^{2}}{m_{A}^{2}}}\mathcal{A}_{S}(m).
  \label{eq:form_factor_S}
\end{equation}

\noindent
Here the \emph S-wave amplitude $\mathcal{A}_{S}(m)$ is considered as a combination 
of a non-resonant background and the $\bar K_{0}^{*}(1430)^{0}$. According to 
the Watson theorem~\cite{km watson}, for the same isospin and angular momentum, 
the phase measured in $K\pi$ elastic scattering and in a decay channel 
are equal in the elastic regime. So the formalism of
the phase of the non-resonant background can 
be taken from the LASS scattering experiment~\cite{lass}.
The total \emph S-wave phase $\delta_{S}(m)$ 
and the amplitude $\mathcal{A}_{S}(m)$ are parameterized in the same way 
as by the BABAR collaboration~\cite{pwa2}:
\begin{eqnarray}
  \rm cot(\delta^{1/2}_{\rm BG})&=&\frac{1}{a_{\rm S,BG}^{1/2}p^{*}}+
  \frac{b_{\rm S,BG}^{1/2}p^{*}}{2},\\
  \rm cot(\delta_{\bar K_{0}^{*}(1430)^{0}})&=&\frac{m^{2}_{\bar K_{0}^{*}(1430)^{0}}-
  m^{2}}{m_{\bar K_{0}^{*}(1430)^{0}}\Gamma_{\bar K_{0}^{*}(1430)^{0}}(m)},\\
  \delta_{S}(m)&=&\delta^{1/2}_{BG}+\delta_{\bar K_{0}^{*}(1430)^{0}},
  \label{eq:phase_S}
\end{eqnarray}

\noindent
where the scattering length $a_{\rm S,BG}^{1/2}$ 
and the effective range $b_{\rm S,BG}^{1/2}$  are determined 
by the PWA fit. $m_{\bar K_{0}^{*}(1430)^{0}}$
is the pole mass of the $\bar K_{0}^{*}(1430)^{0}$.
 $\Gamma_{\bar K_{0}^{*}(1430)^{0}}(m)$ 
is its mass-dependent width, which can be calculated 
using Eq.~(\ref{eq:width}) given the total width
$\Gamma^{0}_{\bar K_{0}^{*}(1430)^{0}}$.

The amplitude $ \mathcal{A}_{S}(m)$ is expressed as

\noindent
 \begin{equation}
 \begin{aligned}
   \mathcal{A}_{S}(m)&=r_{S}P(m)e^{i\delta_{S}(m)},  m<m_{\bar K_{0}^{*}(1430)^{0}};   \\
   \mathcal{A}_{S}(m)&=r_{S}P(m_{\bar K_{0}^{*}(1430)^{0}}) e^{i\delta_{S}(m)} \times  \\
  &\sqrt
  {\frac{(m_{\bar K_{0}^{*}(1430)^{0}}\Gamma^{0}_{\bar K_{0}^{*}(1430)^{0}})^{2}}
  {(m^{2}_{\bar K_{0}^{*}(1430)^{0}}-m^{2})^{2}+(m_{\bar K_{0}^{*}(1430)^{0}}
  \Gamma_{\bar K_{0}^{*}(1430)^{0}}(m))^{2}}}, \\ 
  &m>m_{\bar K_{0}^{*}(1430)^{0}}.
  \label{eq:amplitude_S}
\end{aligned}
\end{equation}

\noindent
Here $P(m) = 1 + x \cdot r_{S}^{(1)}$,  and
$x=\sqrt{\left(\frac{m}{m_{K}+m_{\pi}}\right)^{2}-1}$. 
The dimensionless coefficient 
$r_{S}^{(1)}$ and the relative intensity $r_{S}$ 
are determined by the PWA fit. 

The \emph D-wave related form factors $F_{i2}$ 
are expressed similarly to those of the \emph P-wave:
\begin{eqnarray}
 \label{eq:form_factor_D}
  \mathcal{F}_{12}&=&\frac{m_{D}p_{K\pi}}{3}[(m_{D}^{2}-m^{2}-q^{2})(m_{D}+m)T_{1}(q^{2}) 
  \nonumber \\
  &-&\frac{m_{D}^{2}p_{K\pi}^{2}}{m_{D}+m}T_{2}(q^{2})]\mathcal{A}(m), 
  \nonumber \\
  \mathcal{F}_{22}&=&\sqrt{\frac{2}{3}}m_{D}mqp_{K\pi}(m_{D}+m)T_{1}(q^{2})\mathcal{A}(m), \\
  \mathcal{F}_{32}&=&\sqrt{\frac{2}{3}}\frac{2m_{D}^{2}mqp_{K\pi}^{2}}{m_{D}+m}T_{V}(q^{2})\mathcal{A}(m).
 \nonumber
\end{eqnarray}
 
For the \emph D-wave, we still assume that there are one vector $T_{V}(q^2)$ and 
two axial-vector $T_{1,2}(q^2)$ form factors, 
which behave according to the SPD model. Pole masses 
are assumed to be
 the same as those of the \emph P-wave, and the form factor ratios 
$r_{22}=T_{2}(0)/T_{1}(0)$ and $r_{2V}=T_{V}(0)/T_{1}(0)$ 
at $q^{2}=0$ are expected to be 1~\cite{ak leibovich}.
The amplitude $\mathcal{A}(m)$ is described by the formula in 
Eq.~(\ref{eq:BW}) in the case of $J=2$.

The PWA is performed using an unbinned maximum likelihood fit.
The likelihood expression is

\noindent
\begin{equation}
  L=\prod_{i=1}^{N}\text{PDF}(\xi_{i}, \eta)
   =\prod_{i=1}^{N} \frac{\omega(\xi_{i}, \eta)\epsilon(\xi_{i})}
   {\int d\xi_{i} \omega(\xi_{i}, \eta) \epsilon(\xi_{i})},
\label{eq:l}
\end{equation}

\noindent
where $N$ denotes the number of the events in the PWA.
PDF($\xi, \eta$) is the probability density 
function with arguments
$\xi$ denoting the five kinematic variables characterizing the event,
and $\eta$ denoting the fit parameters. 
$\omega(\xi, \eta)$ and $\epsilon(\xi)$  represent the 
decay intensity (\emph{i.e.}, $\mathcal{I}$ in 
Eq.~(\ref{eq:decay_rate})) and  the acceptance for events of $\xi$.

Omitting the terms independent of the fit parameters
we obtain the negative log-likelihood:

\noindent
\begin{equation} 
  -\ln L=-\sum_{i=1}^{N}\ln\frac{\omega(\xi_{i}, \eta)}{\sigma(\eta)}.
  \label{eq:nll}
\end{equation}

\noindent
The acceptance is taken into account in the term $\sigma(\eta)$,
which is calculated using the PWA signal MC events that pass the event selection~\cite{pwa6}:

\noindent
\begin{equation}
\sigma(\eta)=\int d\xi \omega(\xi, \eta) \epsilon(\xi)
      \propto\frac{1}{N_{\rm selected}}\sum_{k=1}^{N_{\rm selected}}
\frac{\omega(\xi_{k}, \eta)}{\omega(\xi_{k}, \eta_{0})},
\label{eq:mc integral}
\end{equation}

\noindent
where $\eta_{0}$ denotes the set of the 
parameters used to produce the simulated events.

The effect of background in the fit is considered
by subtracting its contribution in the likelihood calculation using 
Eq.~(\ref{eq:nll}):

\noindent
\begin{equation}
  -\ln L_{\rm final}=(-\ln L_{\rm data})-(-\ln L_{\rm bkg}),
  \label{eq:nll_bg}
\end{equation}

\noindent
where $L_{\rm data}$ and $L_{\rm bkg}$ represent the likelihoods of the 
data sample and the background, respectively.  $-\ln L_{\rm final}$ is minimized
to determine the PWA solution. 
$L_{\rm bkg}$ is calculated using the non-signal $D^{+}$ decays
and the combinatorial background, as introduced in Sec.~\ref{sec:select}.

The goodness of the fit is estimated using $\chi^2 / \text{n.d.f.,}$
 where  $\text{n.d.f.}$ denotes the number of degrees of freedom.
 The $\chi^2$ is calculated 
from the difference of the event distribution between data and MC
predicted by the fit in the 
five-dimensional space of  the kinematic
variables $m$, $q^2$, $\cos\theta_{K}$,
$\cos\theta_{e}$ and $\chi$ initially divided into 4, 3, 3, 3 and 3 bins.
The bins are set with different sizes so that they contain 
approximately equal number of signal events.
Each five-dimensional bin is required to contain at least 
10 events, otherwise it is combined with an adjacent bin.
The $\chi^2$ value  is calculated as:

\noindent
\begin{equation}
  \chi^{2}=\sum_{i}^{N_\text{bin}}\frac{(n_{i}^\text{data}-n_{i}^\text{fit})^2}{n_{i}^\text{fit}},
  \label{eq:chi}
\end{equation}

\noindent
where $N_\text{bin}$ is the number of the bins, $n_{i}^\text{data}$ denotes 
the measured content of the $i_\text{th}$ bin, 
and $n_{i}^\text{fit}$ denotes the expected $i_\text{th}$ bin content  
predicted by the fitted PDF.
The $\text{n.d.f.}$ is equal to 
the number of the bins ($N_\text{bin}$) minus the number of the fit 
parameters minus 1.

The structure of the $K\pi$ system is dominated
by the $\bar K^{*}(892)^{0}$. 
As for other possible components, we determine their significances
from the change of -2ln$L$ in the PWA fits with and without contribution of the component,
taking into account the change of the $\text{n.d.f.}$.
The contribution of the \emph S-wave (the $\bar K_{0}^{*}(1430)^{0}$ and the non-resonant part)
is observed with a significance far larger than $10\,\sigma$.
The solution including the $\bar K^{*}(892)^{0}$ and the  \emph S-wave,
with the magnitude and phase of the $\bar K^{*}(892)^{0}$
component fixed at 1 and 0,
is referred to here as ``nominal solution''. 
The contribution from the $\bar K^{*}(1680)^{0}$ is ignored because it is 
suppressed by the small phase space available. We also assume
the contribution from the $\kappa$ to be negligible, 
as follows from the FOCUS  results~\cite{focus621}.
Possible contributions from 
the $\bar K^{*}(1410)^{0}$  and $\bar K_{2}^{*}(1430)^{0}$ are searched.

The fraction of each component can be determined by the ratio of 
the decay intensity of the specific component and that of the total:

\noindent
\begin{equation}
f_{k}=\frac{\int d\xi \omega_{k}(\xi, \eta)}{\int d\xi \omega(\xi, \eta)},
\label{eq:fraction}
\end{equation}

\noindent
where $\omega_{k}(\xi, \eta)$ and $\omega(\xi, \eta)$ denote the decay intensity of 
component $k$ and the total, respectively.

\begin{figure}[htp]  
  \begin{center}
  \includegraphics[width=0.48\linewidth]{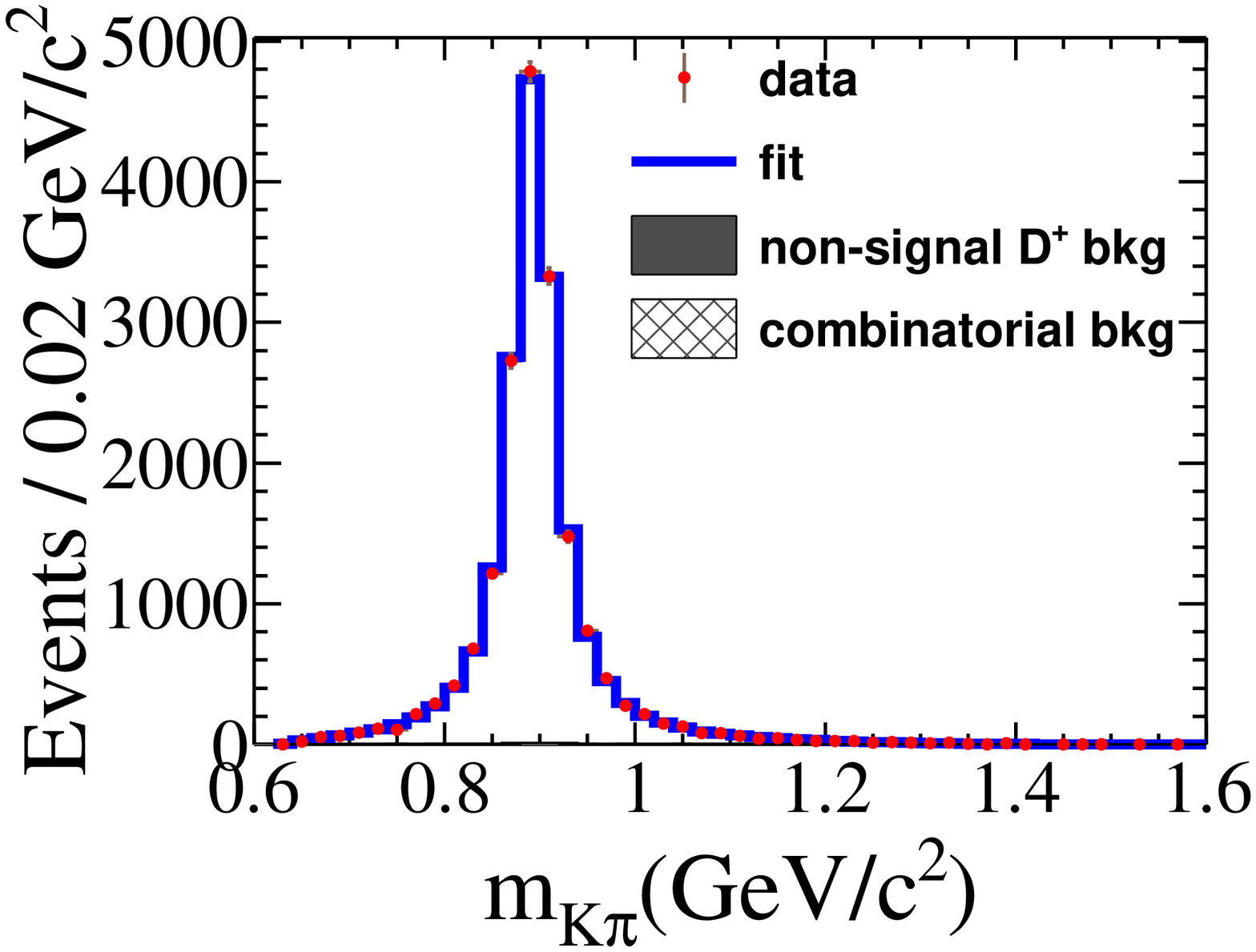}
  \includegraphics[width=0.48\linewidth]{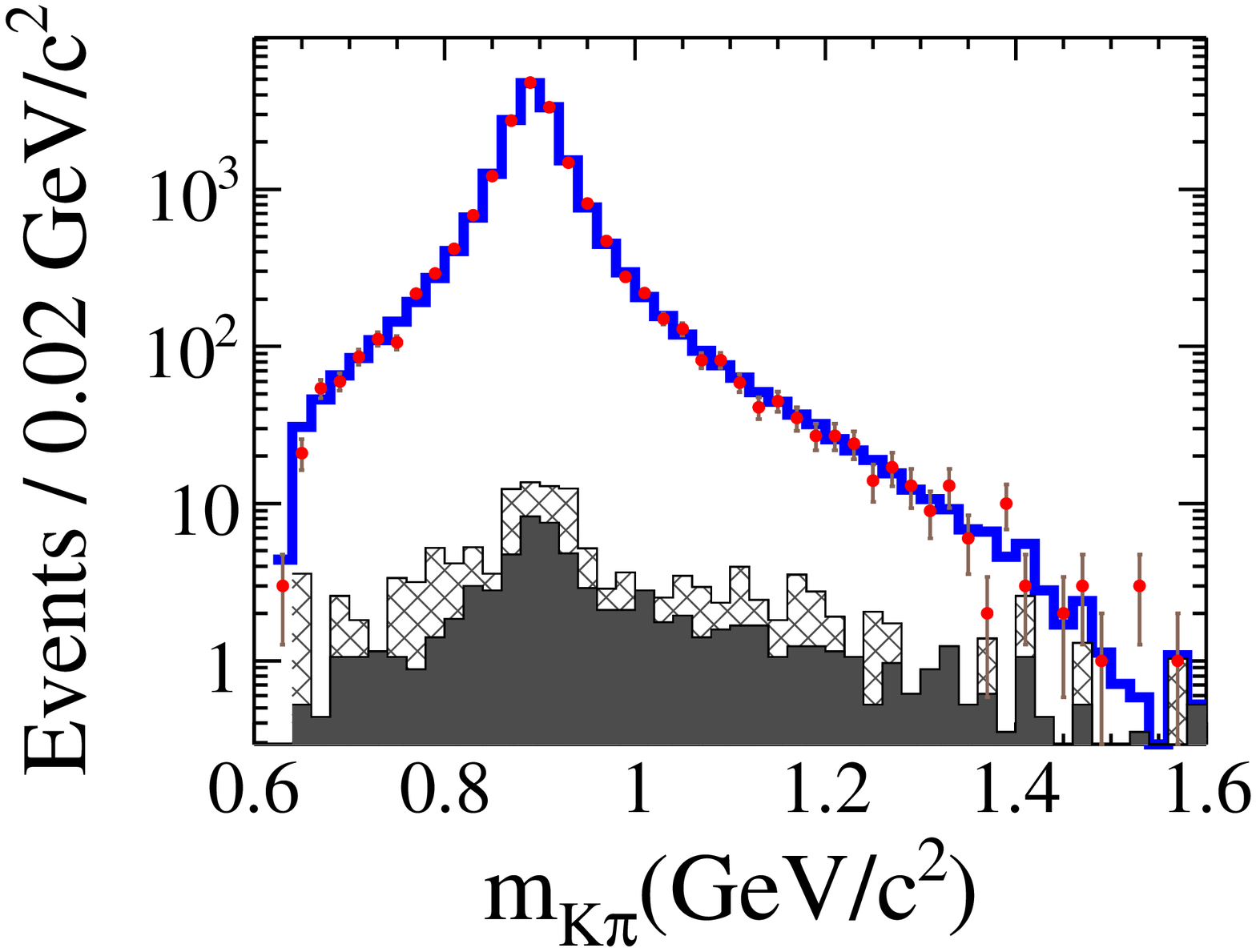} \\
  \includegraphics[width=0.48\linewidth]{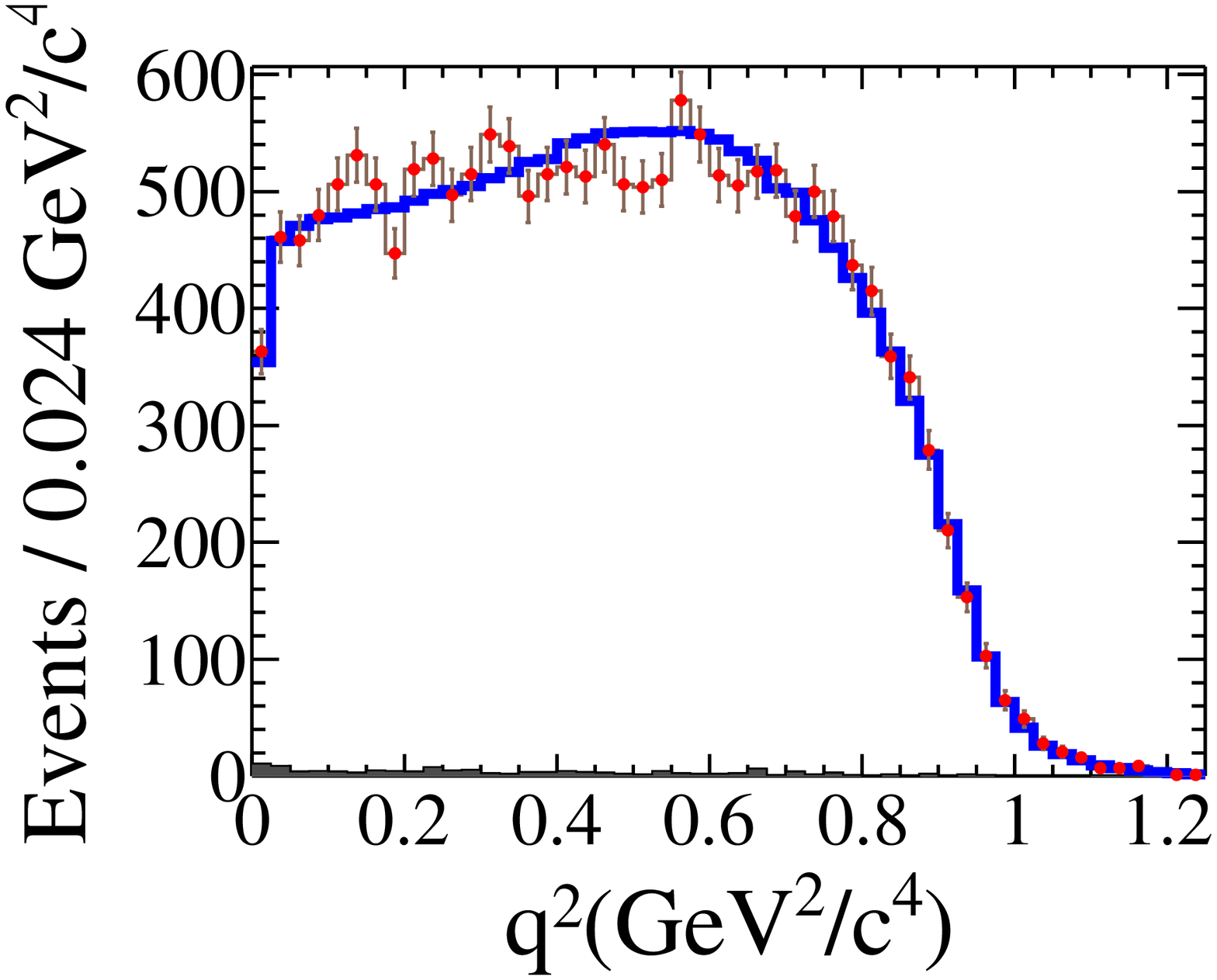} 
  \includegraphics[width=0.48\linewidth]{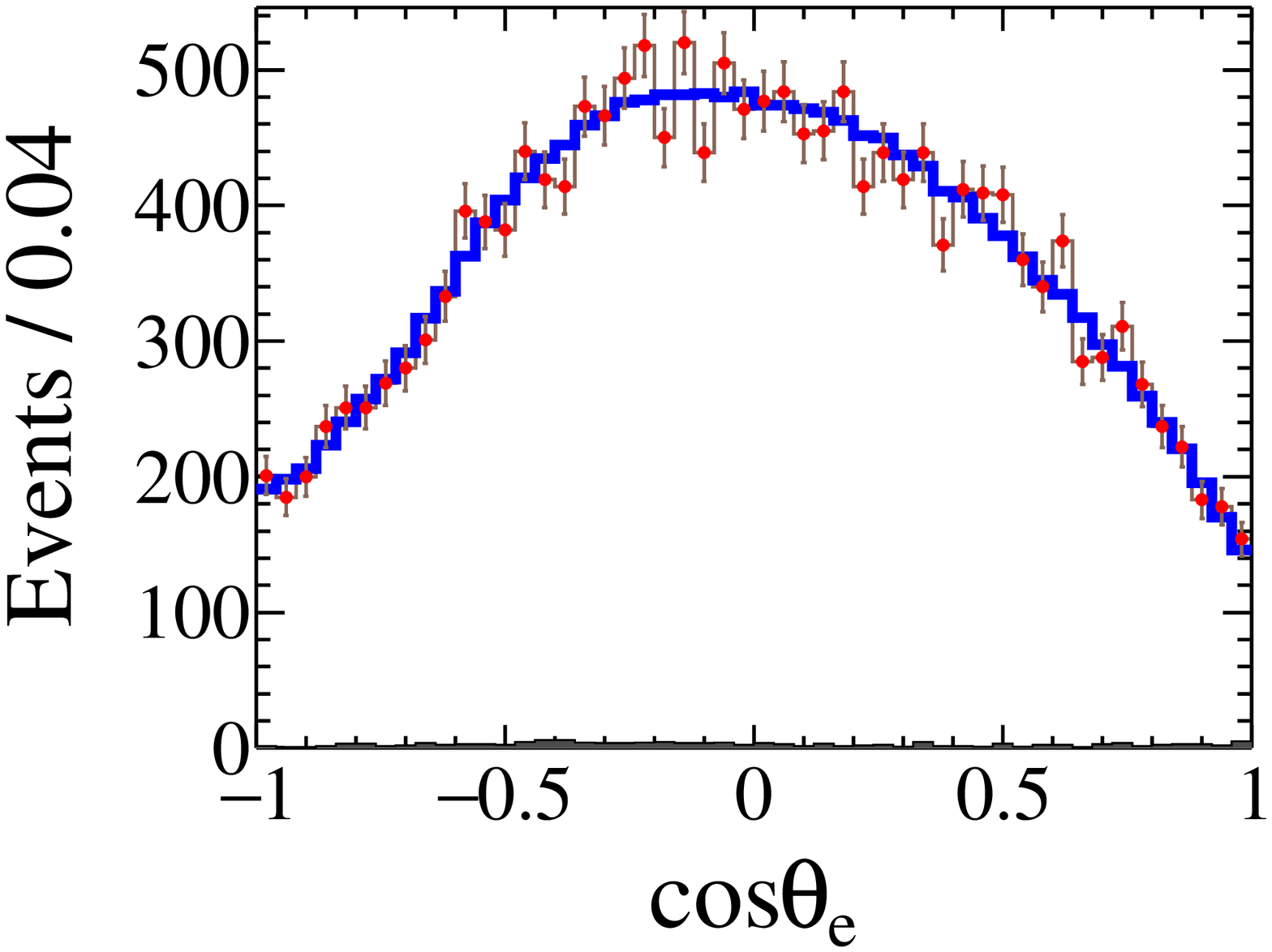} \\
  \includegraphics[width=0.48\linewidth]{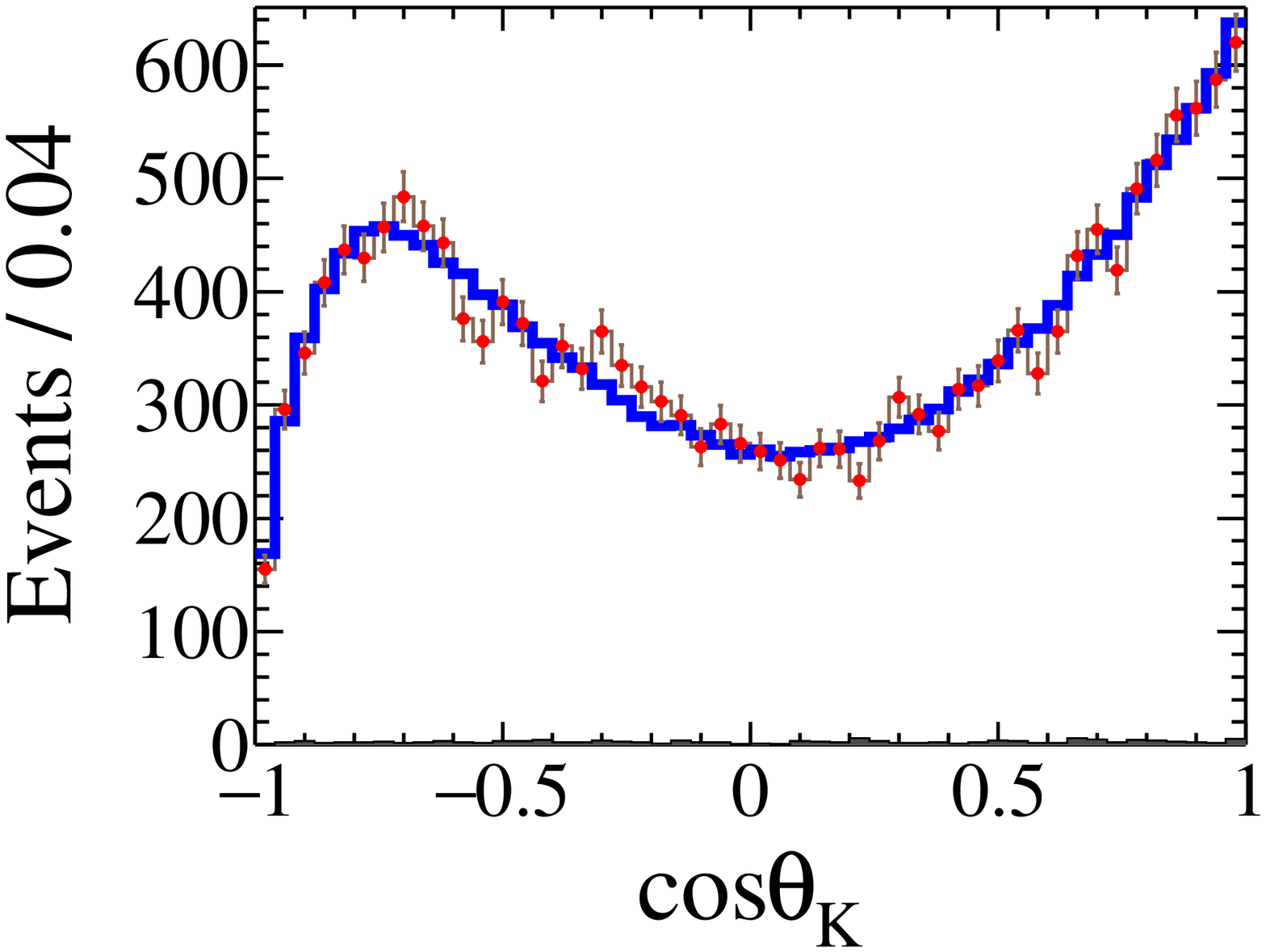}
  \includegraphics[width=0.48\linewidth]{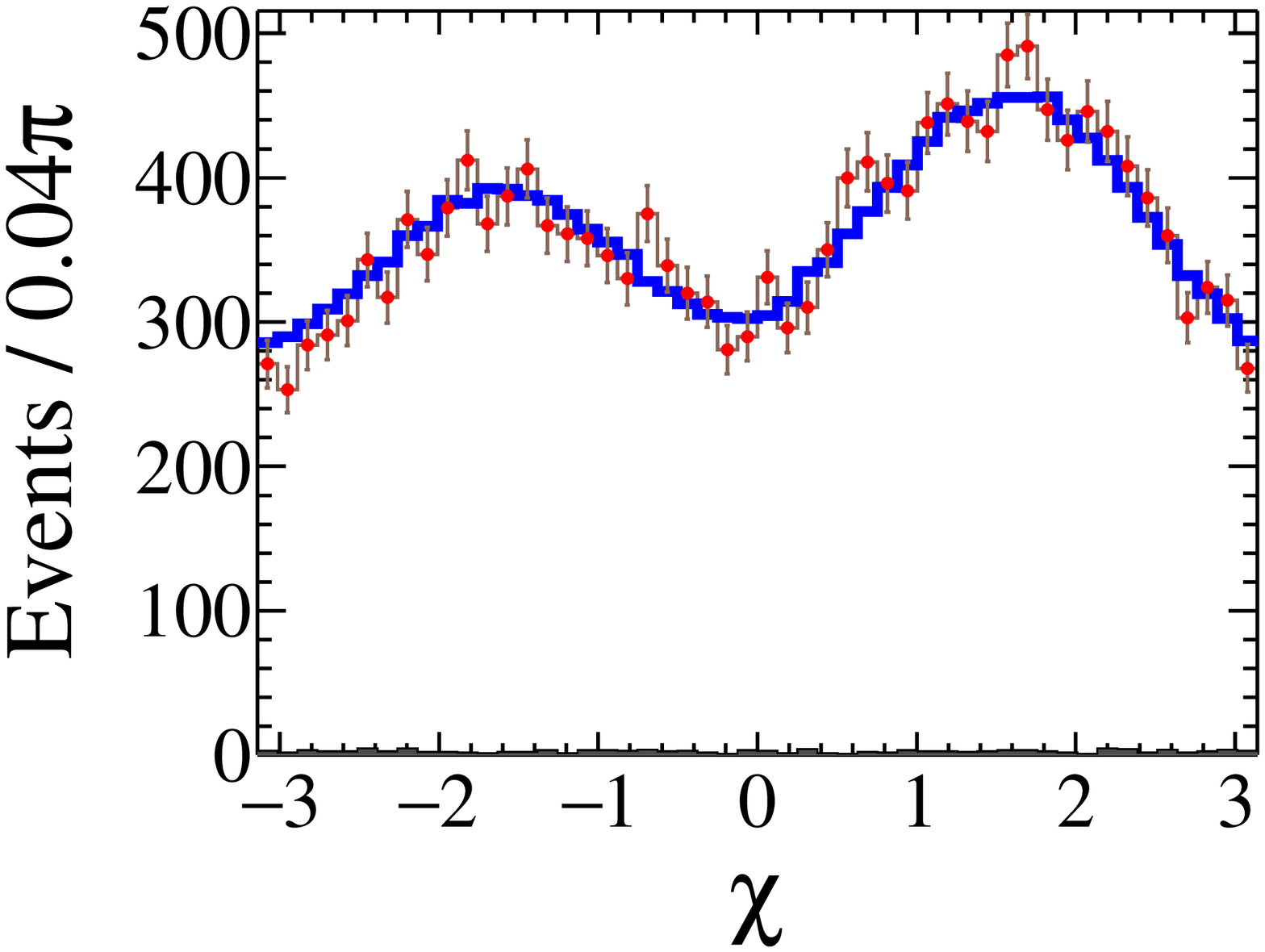}
  \caption{Projections onto each of the kinematic variables, 
  comparing data (dots with error bars) and signal MC 
  determined by PWA solution (solid line),
  assuming that the signal is composed of the \emph S-wave and the $\bar K^{*}(892)^{0}$. 
 The shadowed histogram
  shows the non-signal $D^+$ background estimated from MC simulation 
  and the hatched area shows the combinatorial background
  estimated from the $M_{\rm BC}$ sideband of data.}
  \label{fig:projection_SP}
  \end{center}
\end{figure}

The nominal solution of the PWA fit,
together with the fractions of both components and the goodness of the fit,
are listed in the second column of Table~\ref{tab:pwa}. 
Comparisons of the projections over the five kinematic
variables between data and the PWA solution are illustrated 
in \figref{fig:projection_SP}.

Using the result of $\mathcal{B}(D^{+} \to K^{-} \pi^+ e^+ \nu_e)$
 from Eq.~(\ref{eq:br_total}), 
the branching fractions of both components are calculated to be

\noindent
\begin{equation}
\begin{aligned} 
 &\mathcal{B}(D^{+}\to K^{-} \pi^{+} e^{+}\nu_{e})_{S-\rm wave} 
  = (0.228\pm0.008\pm0.008)\%, \\
 & \mathcal{B}(D^{+}\to K^{-}\pi^{+} e^{+}\nu_e)_{\bar{K}^{*}(892)^{0}} 
  = (3.54\pm0.03\pm0.08)\%,
\label{eq:Br_892_S_val}
\end{aligned} 
\end{equation}

\noindent
where the first errors are statistical and the second 
systematic (described later in this section).

\begin{figure}[htp]
  \begin{center}
  \includegraphics[width=0.7\linewidth]{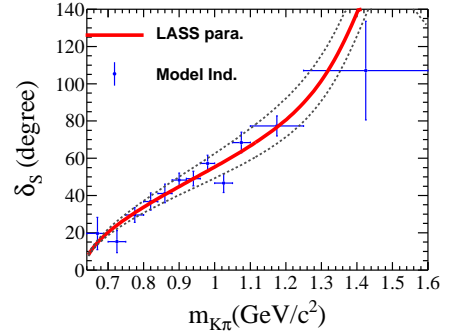}
  \caption{Variation of the \emph S-wave phase versus $m_{K\pi}$,
  assuming that the signal is composed of the
  \emph S-wave and the $\bar K^{*}(892)^{0}$. 
  The points with error bars correspond to the model-independent 
  measurement by fitting data; the solid line 
  corresponds to the result based 
  on the LASS parameterization: $a^{1/2}_{\rm B,SG}=1.94$, 
  $b^{1/2}_{\rm B,SG}=-0.81$; 
  the dotted line shows the
$1\,\sigma$ confidence band
by combining the statistical and systematic errors in quadrature.
}
  \label{fig:pwa_Sphase}
  \end{center}
\end{figure}

The nominal solution is based on the 
$\delta_{S}$ parameterization from Eq.~(\ref{eq:phase_S}).
To test the applicability of this parameterization,
the $m_{K\pi}$ spectrum is divided into 12 bins
and the PWA fit is performed with the phases $\delta_{S}$
in each bin as 12 additional fit parameters (within each bin,
the phase is assumed to be constant).
The measured invariant mass dependence of the phase
is summarized in Table~\ref{tab:pwa_Sphase}.
All other parameters are consistent with those in 
the nominal fit.
Figure~\ref{fig:pwa_Sphase} illustrates the comparison of 
the model-independent measurement with that 
based on the parameterization from Eq.~(\ref{eq:phase_S}).

Possible contributions from the $\bar K^{*}(1410)^{0}$ 
and $\bar K_{2}^{*}(1430)^{0}$
are studied by adding these resonances to the nominal 
solution with the complex coefficients 
$r_{\bar K^{*}(1410)^{0}}e^{i\delta_{\bar K^{*}(1410)^{0}}}$ 
and $r_{\bar K_{2}^{*}(1430)^{0}}e^{i\delta_{\bar K_{2}^{*}(1430)^{0}}}$. 
Due to the scarce population in the high $K\pi$ mass region, 
this analysis is not sensitive to the shapes of these resonances.
Their masses and widths are therefore fixed at the values from PDG. 
They are added to the nominal solution one by one. 
The effective range parameter $b_{\rm S,BG}^{1/2}$ is 
fixed at the result from the nominal solution. 
Based on the isobar model, time reversal symmetry requires the coupling constants 
for the $\bar K^{*}(1410)^{0}$ and $\bar K_{2}^{*}(1430)^{0}$
 to be real, which means that the phases of 
the $\bar K^{*}(1410)^{0}$ and $\bar K_{2}^{*}(1430)^{0}$ 
are only allowed to be zero or $\pi$.

The fit results are summarized in the third 
and fourth columns of Table~\ref{tab:pwa}. 
The contribution from the $\bar K^{*}(1410)^{0}$ is found to 
be consistent with zero when fixing $\delta_{\bar K^{*}(1410)^{0}}$
either at zero or $\pi$,
while the $\bar K_{2}^{*}(1430)^{0}$
has a significance of $4.3\,\sigma$, favoring 
$\delta_{\bar K_{2}^{*}(1430)^{0}}$ at zero. 
The upper limits of their branching fractions
at 90\% confidence level (C.L.) are 
calculated using a Bayesian approach.
They are determined as the branching fraction
below which lies 90\% of the total likelihood integral in
the positive branching fraction domain, assuming a uniform
prior. To take the systematic uncertainty into account,
the likelihood is convolved with a Gaussian function
with a width equal to the systematic uncertainty.
The branching fractions and their upper limits are measured to be

\noindent
\begin{equation}
\begin{aligned} 
\mathcal{B}(D^{+}\to \bar{K}^{*}(1410)^{0} e^{+}\nu_e) 
       &= (0\pm0.009\pm0.008)\%, \\
       &< 0.028\%\,~ (90\% \, \rm C.L.). \\
\mathcal{B}(D^{+}\to \bar{K}^{*}_{2}(1430)^{0} e^{+}\nu_e) 
    &= (0.011\pm0.003\pm0.007)\% ,\\
    &< 0.023\% \,~ (90\%\, \rm C.L.).
\label{eq:eq_14*0}
\end{aligned} 
\end{equation}

\begin{table*}[htp]
  \begin{center}
  \renewcommand{\arraystretch}{1.2}
  \renewcommand{\tabcolsep}{0.03 \linewidth}
  \caption{The PWA solutions with different combinations 
  of \emph{S}(the $\bar K_{0}^{*}(1430)^{0}$  
  and the non-resonant part), 
  \emph{P}($\bar K^{*}(892)^{0}$), 
  \emph{P$^{'}$}($\bar K^{*}(1410)^{0}$) 
  and \emph{D}($\bar K_{2}^{*}(1430)^{0}$)
  components. The first and second uncertainties are 
  statistical and systematic, respectively.}
  \begin{tabular}{p{2.8 cm}    c   c  c }
  \hline
  \hline
  Variable & \emph{S}+\emph{P} & 
  \emph{S}+\emph{P}+\emph{P$^{'}$}  &  
  \emph{S}+\emph{P}+\emph{D}    \\
  \hline  
$r_{S}$(GeV)$^{-1}$ & -11.57$\pm$0.58$\pm$0.46 & -11.57$\pm$0.61$\pm$0.44 & -11.94$\pm$0.58$\pm$0.50  \\
$r_{S}^{(1)}$ & 0.08$\pm$0.05$\pm$0.05 & 0.08$\pm$0.05$\pm$0.05 & 0.03$\pm$0.05$\pm$0.07  \\
$a_{\rm S,BG}^{1/2}$(GeV/$c$)$^{-1}$ & 1.94$\pm$0.21$\pm$0.29 & 1.93$\pm$0.16$\pm$0.50 & 1.84$\pm$0.10$\pm$0.47  \\
$b_{\rm S,BG}^{1/2}$(GeV/$c$)$^{-1}$ & -0.81$\pm$0.82$\pm$1.24 & -0.81 fixed & -0.81 fixed  \\
$m_{\bar K^{*}(892)^{0}}$(MeV/$c^{2}$) & 894.60$\pm$0.25$\pm$0.08 & 894.61$\pm$0.35$\pm$0.12 & 894.68$\pm$0.25$\pm$0.05  \\
$\Gamma_{\bar K^{*}(892)^{0}}^{0}$ (MeV/$c^{2}$) & 46.42$\pm$0.56$\pm$0.15 & 46.44$\pm$0.70$\pm$0.26 & 46.53$\pm$0.56$\pm$0.31  \\
$r_{\rm BW}$ (GeV/$c$)$^{-1}$ & 3.07$\pm$0.26$\pm$0.11 & 3.05$\pm$0.61$\pm$0.30 & 3.01$\pm$0.26$\pm$0.22  \\
$m_{V}$ (GeV/$c^{2}$) & 1.81$^{+0.25}_{-0.17}\pm$0.02 & 1.81$^{+0.25}_{-0.17}\pm$0.02 & 1.80$^{+0.24}_{-0.16}\pm$0.05 \\
$m_{A}$ (GeV/$c^{2}$) & 2.61$^{+0.22}_{-0.17}\pm$0.03 & 2.60$^{+0.22}_{-0.17}\pm$0.03 & 2.60$^{+0.21}_{-0.17}\pm$0.04  \\
$r_{V}$ & 1.411$\pm$0.058$\pm$0.007 & 1.410$\pm$0.057$\pm$0.006 & 1.406$\pm$0.058$\pm$0.022  \\
$r_{2}$ & 0.788$\pm$0.042$\pm$0.008 & 0.788$\pm$0.041$\pm$0.008 & 0.784$\pm$0.041$\pm$0.024  \\
$r_{\bar K^{*}(1410)^{0}}$ & &  0.00$\pm$0.40$\pm$0.04  \\
$\delta_{\bar K^{*}(1410)^{0}}$(degree) & & 0 fixed  \\
$r_{\bar K_{2}^{*}(1430)^{0}}$(GeV)$^{-4}$ & & & 11.22$\pm$1.89$\pm$4.10 \\
$\delta_{\bar K_{2}^{*}(1430)^{0}}$(degree)& & & 0 fixed  \\
\hline
$f_{S}$(\%)   &   6.05$\pm$0.22$\pm$0.18  & 6.06$\pm$0.24$\pm$0.18  
              &  5.90$\pm$0.23$\pm$0.20  
\\
$f_{\bar K^{*}(892)^{0}}$(\%)   & 93.93$\pm$0.22$\pm$0.18  &   93.91$\pm$0.24$\pm$0.18  
                       &  94.00$\pm$0.23$\pm$0.16 \\
$f_{\bar K^{*}(1410)^{0}}$(\%)  & & 0$\pm$0.010$\pm$0.009 \\
                   
$f_{\bar K_{2}^{*}(1430)^{0}}$(\%) & & & 0.094$\pm$0.030$\pm$0.061 \\
                      
   \hline 
  $\chi^{2} / n.d.f.$            &    292.7/291  &  292.7/291 &  292.7/292  \\
  \hline  
  \hline
  \end{tabular}
  \label{tab:pwa}
  \end{center}
\end{table*}

We also try to add both  the $\bar K^{*}(1410)^{0}$  
and $\bar K_{2}^{*}(1430)^{0}$ to 
the fit, obtaining  results that are quite close to the solution
in the fourth column of Table~\ref{tab:pwa}.
This suggests that the $\bar K^{*}(1410)^{0}$ contribution
can be neglected.

In the PWA fit, only the ratios of the transition
form factors $r_{V}$ and $r_{2}$ are measured. Given the result of 
$\mathcal{B}$($D^{+}\to \bar{K}^{*}(892)^{0} e^{+} \nu_{e}$)
from Eq.~(\ref{eq:Br_892_S_val}), we can calculate the 
$A_{1}(0)$ value and thus obtain the absolute values 
of the form factors, which can be compared with 
the lattice QCD determinations.

The value of $A_{1}(0)$ is calculated by comparing 
the absolute branching fraction and the integration 
of the differential decay rate given in Eq.~(\ref{eq:decay_rate})
over the five-dimensional space
for the $D^{+}\to \bar{K}^{*}(892)^{0} e^{+} \nu_{e}$ process.
Restricting Eq.~(\ref{eq:decay_rate}) to the $\bar{K}^{*}(892)^{0}$
contribution only and integrating it
over the three angles, we obtain 

\noindent
\begin{equation}
\frac{d\Gamma}{dq^{2}dm^{2}}=\frac{1}{3}
\frac{G_{F}^{2}|V_{cs}|^{2}}
{(4\pi)^{5}m_{D}^{2}}\beta p_{K\pi}
\bigg[\frac{2}{3}\big\{|\mathcal{F}_{11}|^{2}+|\mathcal{F}_{21}|^{2}+|\mathcal{F}_{31}|^{2}\big\}\bigg].
\label{eq:A1_gamma}
\end{equation}

Assuming that $\bar{K}^{*}(892)^{0}$ has an infinitesimal width and 
 a single pole mass of 894.60 MeV/$c^{2}$, and integrating
 Eq.~(\ref{eq:A1_gamma}) over $q^{2}$, we find

\noindent
\begin{eqnarray}
\Gamma &=& \frac{G_{F}^{2}|V_{cs}|^{2}}{96\pi^{3}m_{D}^{2}}
   \frac{2}{3}|A_{1}(0)|^{2}\mathbb{X}  \\
&\equiv& \frac{\hbar \mathcal{B}(D^{+}\to \bar{K}^{*}(892)^{0} e^{+} \nu_{e})
    \mathcal{B}(\bar{K}^{*}(892)^{0} \to K^{-} \pi^{+})}{\tau_{D^{+}}} \nonumber
\end{eqnarray}
with
\begin{eqnarray}
\mathbb{X}=\int_{0}^{q^{2}_{\rm max}}p_{K \pi}q^{2}
     \frac{|H_{0}|^{2}+|H_{+}|^{2}+|H_{-}|^{2}}{|A_{1}(0)|^{2}}dq^{2}. \nonumber
\end{eqnarray}

\noindent
Here $\hbar$ is the reduced Planck constant and $\tau_{D^{+}}$
is the lifetime of $D^{+}$ meson.
The integral $\mathbb{X}$ is evaluated using 
$r_{2}$, $r_{V}$, $m_{V}$ and $m_{A}$ from the PWA solution.
Using the values $\tau_{D^{+}}=(10.40\pm0.07)\times 10^{-13}$s 
and $|V_{cs}|=0.986\pm0.016$
from PDG, one gets 

\noindent
\begin{equation}
A_{1}(0)=0.589\pm0.010\pm0.012.
\label{eq:A1_0_fixW}
\end{equation}

This result is more than one standard deviation lower than that in Ref.~\cite{pwa2}.
The difference can mostly be explained by the lower value of 
$\mathcal{B}$($D^{+}\to \bar{K}^{*}(892)^{0} e^{+} \nu_{e}$)
in Eq.~(\ref{eq:Br_892_S_val}) and by the renewed measurement 
of $|V_{cs}|$ in PDG.

If instead of approximating the $\bar{K}^{*}(892)^{0}$ mass distribution 
as a delta-function, we use the fitted mass distribution of the resonance to integrate 
the differential decay rate over $q^2$ and $m^2$, the result becomes

\noindent
\begin{equation}
A_{1}(0)|_{q^2, m^2}=0.619\pm0.011\pm0.013,
\label{eq:A1_0_floatW}
\end{equation}

\noindent
where the integration for $m^2$ is performed  over the mass range
0.6$<$$m_{K\pi}$$<$1.6 GeV/$c^{2}$.
We do not observe the large difference between $A_{1}(0)$
and $A_{1}(0)|_{q^2, m^2}$ reported in Ref.~\cite{pwa2}.

In PWA, the systematic uncertainty of each parameter is 
defined as the difference between the fit result 
in the nominal condition and that obtained after some 
condition is varied corresponding to one source of uncertainty.
Systematic uncertainties of the  nominal solution
are summarized in Table~\ref{tab:pwasyst}.  The uncertainty
due to the background fraction is
estimated by varying the background fraction by $1\,\sigma$
in the same way as when estimating this uncertainty 
 in branching fraction measurement in Sec.~\ref{sec:BF}.
Uncertainties due to the assumed shapes of the backgrounds are considered 
separately for the combinatorial background and 
the non-signal $D^{+}$ decays. The former is estimated by
varying the $M_{\rm BC}$ sideband,
while for the latter only the uncertainty 
from $D^{+}\to K^{-} \pi^{+} \pi^{+} \pi^{0}$ is considered, 
which is estimated by comparing the difference 
between two extreme cases: phase space process and 
$D^{+}\to \bar{K}^{*}(892)^{0} \rho^{+}$. 
 The uncertainty due to the shape of the other 
non-signal $D^{+}$ decays can be neglected. 
The uncertainty arising from the fixed mass and width 
of  the $\bar K_{0}^{*}(1430)^{0}$ is considered by 
varying their values by $1\,\sigma$ according 
to PDG. To estimate the uncertainty caused by the additional resonances,
we compare different solutions 
in Table~\ref{tab:pwa} and take the largest differences
between them as systematic uncertainties.
$b^{1/2}_{S,BG}$ has been fixed 
in solutions with the $\bar{K}^{*}(1410)^{0}$ or 
$\bar{K}_{2}^{*}(1430)^{0}$ component considered.
We then allow it to be a free  parameter in the fits 
and the largest variation of $b^{1/2}_{S,BG}$ is taken as the uncertainty.
The uncertainty associated with the efficiency correction 
of tracking and particle identification is obtained by 
varying the correction factor by $1\,\sigma$. 
The possible  uncertainty due to the fit procedure is studied with 
500 fully reconstructed data-sized signal MC samples 
generated according to the PWA result.
The input-output check shows that  biases of all the fit parameters 
are negligible.
Assuming that all the  uncertainties described above are independent of 
each other, we add them in quadrature to obtain the total.
In a similar way, systematic uncertainties on the \emph S-wave phase $\delta_{S}$
are estimated and presented in Table~\ref{tab:pwa_Sphase}.

\begin{table*}[htp]
  \begin{center}
  \renewcommand{\arraystretch}{1.1}
  \renewcommand{\tabcolsep}{0.01\textwidth}
  \caption{Systematic uncertainties of the PWA nominal  solution
arsing from: 
  (I) background fraction,
  (II) background shape, 
  (III) the $\bar K_{0}^{*}(1430)^{0}$ mass and width,
  (IV) additional resonances,
  (V) tracking efficiency correction,
  (VI) PID efficiency correction.}
  \begin{tabular}{p{4cm}| c c c c c c c c c}
     \hline
     \hline
           Variable                  &   I      &   II   &    III   &  IV   &  V &  VI & total \\ 
     \hline
$\Delta r_{S}$(GeV)$^{-1}$   &  0.03  &  0.26  &  0.10  &  0.37  &  0.01  &  0.01  &  0.46  \\  
$\Delta r_{S}^{(1)}$   &  0.00  &  0.02  &  0.01  &  0.05  &  0.00  &  0.00  &  0.05  \\  
$\Delta a_{\rm S,BG}^{1/2}$(GeV/$c$)$^{-1}$   &  0.01  &  0.04  &  0.27  &  0.10  &  0.01  &  0.00  &  0.29  \\  
$\Delta b_{\rm S,BG}^{1/2}$(GeV/$c$)$^{-1}$   &  0.03  &  0.21  &  1.20  &  0.23  &  0.02  &  0.00  &  1.24  \\  
$\Delta m_{\bar K^{*}(892)^{0}}$(MeV/$c^{2}$)   &  0.00  &  0.02  &  0.00  &  0.07  &  0.00  &  0.00  &  0.08  \\  
$\Delta \Gamma_{\bar K^{*}(892)^{0}}^{0}$ (MeV/$c^{2}$)   &  0.01  &  0.10  &  0.02  &  0.11  &  0.00  &  0.00  &  0.15  \\  
$\Delta r_{\rm BW}$ (GeV/$c$)$^{-1}$   &  0.00  &  0.09  &  0.02  &  0.06  &  0.00  &  0.00  &  0.11  \\  
$\Delta m_{V}$ (GeV/$c^{2}$)   &  0.00  &  0.01  &  0.00  &  0.02  &  0.01  &  0.00  &  0.02  \\  
$\Delta m_{A}$ (GeV/$c^{2}$)   &  0.00  &  0.02  &  0.00  &  0.01  &  0.01  &  0.00  &  0.03  \\  
$\Delta r_{V}$   &  0.001  &  0.004  &  0.001  &  0.005  &  0.001  &  0.001  &  0.007  \\  
$\Delta r_{2}$   &  0.000  &  0.005  &  0.001  &  0.004  &  0.005  &  0.000  &  0.008  \\ 
     \hline 
     \hline 
  \end{tabular}
  \label{tab:pwasyst}
  \end{center}
\end{table*}

\begin{table*}[htp]
  \begin{center}
  \renewcommand{\arraystretch}{1.1}
  \renewcommand{\tabcolsep}{0.01\textwidth}
  \caption{The \emph S-wave phase $\delta_{S}$ measured in the 12 $m_{K\pi}$ bins 
  with statistical and systematic uncertainties.
  The systematic uncertainties include: 
  (I) background fraction,
  (II) background shape, 
  (III) the $\bar K_{0}^{*}(1430)^{0}$ mass and width, 
  (IV) additional resonances,
  (V) tracking efficiency correction,
  (VI) PID efficiency correction.}
  \begin{tabular}{p{2.2cm}| c c c c c c c c c}
     \hline
     \hline
     $m_{K\pi}$ bin      &  Value & Statistical  & \multicolumn{7}{c}{Systematic} \\
     \hline
           (GeV/$c^2$)   &  (degree) & (degree)  &   I      &   II   &    III   &  IV   &  V &  VI & total \\ 
     \hline
 0.60 - 0.70 &  19.63 & 8.58  & 0.08  &  0.42  &  1.10  &  0.52  &  0.19  &  0.10  &  1.31 \\
 0.70 - 0.75 &  15.22 & 5.51  & 0.02  &  2.20  &  0.05  &  0.09  &  0.02  &  0.01  &  2.20  \\  
 0.75 - 0.80 &  29.55 & 3.93  & 0.16  &  0.21  &  0.12  &  0.50  &  0.10  &  0.10  &  0.60  \\  
 0.80 - 0.84 &  36.74 & 4.61  & 0.00  &  0.25  &  0.23  &  0.27  &  0.04  &  0.04  &  0.44  \\  
 0.84 - 0.88 &  41.10 & 4.96  & 0.03  &  0.31  &  0.23  &  0.70  &  0.06  &  0.06  &  0.80  \\  
 0.88 - 0.92 &  48.28 & 3.71  &  0.04  &  0.22  &  0.13  &  0.46  &  0.04  &  0.04  &  0.53  \\  
 0.92 - 0.96 &  49.06 & 3.76  &  0.03  &  0.54  &  0.12  &  1.10  &  0.01  &  0.01  &  1.23  \\  
 0.96 - 1.00 &  57.27 & 4.15  &  0.04  &  0.28  &  0.19  &  1.30  &  0.05  &  0.05  &  1.35  \\  
 1.00 - 1.05 &  46.63 & 4.47  &  0.01  &  0.25  &  0.34  &  2.30  &  0.18  &  0.18  &  2.35  \\  
 1.05 - 1.10 &  68.46 & 5.01  & 0.01  &  1.10  &  0.18  &  2.10  &  0.03  &  0.03  &  2.38  \\  
 1.10 - 1.25 &  77.32 & 4.34  & 0.18  &  1.20  &  1.30  &  2.80  &  0.13  &  0.12  &  3.32  \\  
 1.25 - 1.60 &  107.08 & 11.24  &  0.97  &  10.00  &  9.50  &  20.00  &  1.10  &  1.10  &  24.36  \\ 
     \hline 
     \hline 
  \end{tabular}
  \label{tab:pwa_Sphase}
  \end{center}
\end{table*}

\section{Determination of helicity basis form factors}

In the $K^*$-dominated region, the contribution of non-$\bar{K}^{*}(892)^{0}$ resonances
is negligible and the decay intensity 
 can be parameterized
by helicity basis form factors 
$H_{\pm, 0}(q^2,m^2)$ describing the decay into the
$\bar{K}^{*}(892)^{0}$ vector,
and
by an additional form factor $h_0(q^2,m^2)$ describing the non-resonant
\emph S-wave contribution.
This allows us to transform the matrix element
$\mathcal{I}$ in Eq.~\eqref{eq:decay_rate} into a simplified form \cite{Link:2005dp}.
By performing an integration over the acoplanarity angle $\chi$ and neglecting the terms suppressed
 by the factor $m_e^{2}/q^{2}$, one obtains

 \begin{equation}
 \begin{aligned}
  &\int{\mathcal{I} \,d\chi} = 
  \frac{q^2-m_e^2}{8} \times \\[0cm]
   &   \left\{ \begin{array}{c}
            \left(\left(1 + \cos{\theta_{e}}\right) 
\sin{\theta_{K}}\right)^2
|H_+(q^2,m^2)|^2 |A_{{K}^{*}}(m)|^2 \\
            + \left(\left(1 - \cos{\theta_{e}}\right) 
\sin{\theta_{K}}\right)^2
|H_-(q^2,m^2)|^2 |A_{{K}^{*}}(m)|^2 \\
            + \left(2 \sin{\theta_{e}} \cos{\theta_{K}} \right)^2
|H_0(q^2,m^2)|^2 |A_{{K}^{*}}(m)|^2 \\
            + \underline{8 \sin^2{\theta_{e}} \cos{\theta_{K}}
H_0(q^2,m^2) h_0(q^2,m^2) } \times \\
\underline{Re\{A_S e^{-i\delta_S} A_{{K}^{*}}(m) \}} \\
            + \underline{4 \sin^2{\theta_{e}} A_S^2
|h_0(q^2,m^2)|^2}
      \end{array} \right\}.
  \label{eq:intensity}
  \end{aligned}
\end{equation}
 
\noindent
Here $A_{{K}^{*}}(m)$ denotes the $\bar{K}^{*}(892)^{0}$ amplitude:

\noindent
 \begin{equation}
   A_{{K}^{*}}(m) =
  \frac{   \sqrt{m_0 \Gamma_0}  \left(\frac{ p^*(m)}  {p^*(m_0)}
\right) }
       {        m^2 - m_0^2 + i m_0 \Gamma_0
                \left(\frac{ p^*(m)}  {p^*(m_0)}  \right)^3 },
  \label{eq:bw}
 \end{equation}

\noindent
where  $m_0$ and $\Gamma_0$ are the mass and the width of $\bar{K}^{*}(892)^{0}$
 with their values taken from the second column of Table~\ref{tab:pwa}.

The  underlined terms in Eq.~\eqref{eq:intensity} represent
the non-resonant \emph S-wave contribution which was
described for the first time in Ref.~\cite{Link:2002ev}.
 The mass and $q^2$ dependence of the non-resonant S-wave
amplitude is parameterized as $h_0(q^2,m^2) A_S(m) e^{i \delta_S(m)}$, where the form factor
$h_0(q^2,m^2)$ is not assumed to be the same as $H_0(q^2,m^2)$.
Generally, both the amplitude modulus $A_S(m)$ and the phase $\delta_S(m)$ are mass dependent. 
However in this section, $A_S(m)$ and $\delta_S(m)$ are both assumed
to be constant throughout the ${K}^{*}$-dominated mass region.
The value of $\delta_S = 39^\circ$ 
is taken from Ref.~\cite{Link:2002wg}.

%

The helicity basis form-factor products 
$|H_+(q^2,m^2)|^2$, $|H_-(q^2,m^2)|^2$, $|H_0(q^2,m^2)|^2$,
$A_{S}H_0(q^2,m^2) h_0(q^2,m^2)$, $A_{S}^{2}h_{0}^{2}(q^2,m^2)$
in Eq.~(\ref{eq:intensity}), which we denote with $\alpha=\{+,-,0,I,S\}$ correspondingly,
can be extracted from the angular distributions in Eq.~(\ref{eq:intensity}) 
in a model-independent way using the projective
weighting technique, which was introduced in 
Ref.~\cite{Link:2005dp}.

In general, the form-factor products are functions of $q^2$ and $m^2$. 
However, in this work we measure the average values  over the
relatively narrow $K^{*}$-dominated region. 
Taking $|H_+(q^2,m^2)|^2$ for example,

\noindent
\begin{equation}
|H_+(q^2)|^2 = \frac
{\int |H_+(q^2,m^2)|^2
F(q^2, m^2)  |A_{{K}^{*}}(m)|^2 dm^2}
{\int F(q^2, m^2)  |A_{{K}^{*}}(m)|^2 dm^2},
\end{equation}

\noindent
where the integration is performed   over the mass range
0.8$<$$m$$<$1.0 GeV/$c^{2}$. The kinematic factor $F(q^2, m^2)$ is defined as

 \noindent
\begin{equation}
F(q^2, m^2) =  \frac{(q^2-m_e^2) p_{K\pi} p^{*}}{m q},
\end{equation}

\noindent
where $p_{K\pi}$ and  $p^{*}$ are defined in Sec.~\ref{sec:pwa}.
Similarly, this averaging procedure is also performed for the other form-factor 
products.

 To obtain the form-factor product dependence on $q^2$,
 we divide the $q^2$ range $0<q^2<1.0 \;\text{GeV}^2/c^4$
into 10 equal bins. The form-factor products are to be calculated in each $q^{2}$ bin independently.
 For events in a given $q^2$ bin, we consider 100 two-dimensional
$\Delta\cos\theta_K\times\Delta\cos\theta_e$ angular bins: 
10 equal-size bins in $\cos\theta_K$ times 10 equal-size bins
in $\cos\theta_e$. Each event  is 
assigned a weight to project
out the given form-factor product depending on the angular bin it 
is reconstructed in.

Such a weighting is equivalent to calculating a scalar product
$\vec{P}_\alpha \cdot \vec{D}$. Here  $\vec{D}=\{n_{1} \,  n_{2} ... \, n_{100}\}$
 is a data vector of the observed angular bin populations
whose $j$th component is the number of data events $n_j$ in the $j$th angular bin, $j=1,2...100$.
$\vec{P}_{\alpha}$ is a projection vector for the form factor product $\alpha$,
whose components serve as weights
 applied to the events in a given angular bin. 
Calculating the scalar product $\vec{P}_{\alpha} \cdot \vec{D}$
is equivalent to  weighting events in the first angular bin
by $\left[\vec{P}_{\alpha}\right]_1$,
in the second bin by $\left[\vec{P}_{\alpha}\right]_2$, etc.:

\begin{equation}
 \vec{P}_{\alpha} \cdot \vec{D} = \left[\vec{P}_{\alpha}\right]_1 n_1 + 
\left[\vec{P}_{\alpha}\right]_2 n_2 + \cdots + \left[\vec{P}_{\alpha}\right]_{100} 
n_{100}.
\end{equation}

The weight vector $\vec{P}_{\alpha}$ and the scalar product $\vec{P}_\alpha \cdot \vec{D}$
can be calculated following the idea described below.
Firstly, the data vector  $\vec{D}$ can be written as a sum of contributions from the terms
related to the individual form-factor products in Eq.~\eqref{eq:intensity}:

\noindent
\begin{equation}
\begin{aligned}
 \vec{D} = 
  &  f_+  \vec{m}_{+}+ f_-  \vec{m}_{-} + f_0  \vec{m}_{0} + 
   f_I  \vec{m}_{I}  + f_S   \vec{m}_{S} \\
  =&   \sum_{\alpha}  f_{\alpha} \vec{m}_{\alpha}.
  \label{eq:d_sum}
\end{aligned}
\end{equation}

\noindent
Here the vectors $\vec{m}_\alpha$ represent the angular distributions of the contributions
from the individual form-factor product components of Eq.~\eqref{eq:intensity}
into $\vec{D}$. They are obtained based on  MC simulation which will be discussed later.
The coefficients $f_{\alpha}$ represent the relative ratio of the individual contributions,
which are proportional to the corresponding form-factor products.

If we define a $5 \times  100$ matrix $M$ as

\noindent
\begin{equation}
 M =
 \begin{pmatrix}
   \vec{m}_+ &  \vec{m}_- & \vec{m}_0 & \vec{m}_{I} & \vec{m}_{S}
 \end{pmatrix}^{T}
 \label{eq:m_matrix},
\end{equation}

\noindent
Eq.~\eqref{eq:d_sum} can be transformed into

\noindent
\begin{equation}
 \begin{pmatrix}
   \vec{m}_+ \cdot \vec{D} \\
   \vec{m}_- \cdot \vec{D} \\
   \vec{m}_0 \cdot \vec{D} \\
   \vec{m}_{I} \cdot \vec{D} \\
   \vec{m}_{S} \cdot \vec{D}
 \end{pmatrix}
 =
 M M^T
 \begin{pmatrix}
  f_+ \\
  f_- \\
  f_0 \\
  f_{I} \\
  f_{S}
 \end{pmatrix}.
\label{eq:component_matrix}
\end{equation}

\noindent
The solution of Eq.~\eqref{eq:component_matrix} is

\noindent
\begin{equation}
 \begin{pmatrix}
 f_+ &  f_- & f_0 & f_{I} &  f_{S}
 \end{pmatrix}^{T}
 =
 P \vec{D},
 \label{eq:fffinal}
\end{equation}

\noindent
with the weight matrix $P$ defined by

\noindent
\begin{equation}
P=
 \begin{pmatrix}
   \vec{P}_+ & \vec{P}_- &  \vec{P}_0 & \vec{P}_{I} &  \vec{P}_{S}
 \end{pmatrix}^{T}  =
 \left( M M^T \right)^{-1} M,
 \label{eq:p_matrix}
\end{equation}

\noindent
whose component  $[\vec{P}_{\alpha}]_{k}$ is used as the weight
for the construction of the form-factor product $\alpha$
in the $k_{th}$ angular bin.

The matrix $M$ is obtained by weighting the PHSP signal MC. 
The simulated events pass the usual
procedure of detector simulation and event selection,
allowing correction for the biases due to the finite
detector resolution and selection efficiency.
Each of the $\vec{m}_{\alpha}$ vectors is calculated by weighing
the PHSP sample so that the resulting data reproduces
the distribution of Eq.~\eqref{eq:intensity} with 
the form-factor product $\alpha$ set at 1
and all the others  being equal to 0. 
 For a given event of  $\theta_{e}$, $\theta_{K}$, $m^2$ and $q^2$,
 the following weights are assigned to calculate the corresponding $\vec{m}_{\alpha}$ vector:

\noindent
 \begin{equation}
 \begin{array}{l}
   \omega_+ = F(q^2,{m^2}) {|A_{{K}^{*}}(m)|^2}  
\left(\left(1 + \cos{\theta_{e}}\right) \sin{\theta_{K}}\right)^2,  \\
   \omega_- = F(q^2,{m^2}) {|A_{{K}^{*}}(m)|^2}  
\left(\left(1 - \cos{\theta_{e}}\right) \sin{\theta_{K}}\right)^2, \\
   \omega_0 = F(q^2,{m^2}) {|A_{{K}^{*}}(m)|^2}
\left(2 \sin{\theta_{e}} \cos{\theta_{K}} \right)^2, \\
   \omega_I = 8~F(q^2,{m^2}) {Re\{e^{-i\delta_{S}}
A_{{K}^{*}}(m) \} }  \sin^2{\theta_{e}} \cos{\theta_{K}}, \\
   \omega_S = 4~F(q^2,{m^2}) \sin^2{\theta_{e}}.
 \end{array}
 \label{eq:pwt_m_alpha_weights}
 \end{equation}

Given the matrix $M$ determined by MC simulation, the weight matrix
$P$ can be calculated using Eq.~(\ref{eq:p_matrix})
and the form-factor products can be obtained 
by applying $P$ to the data vector $\vec{D}$
according to Eq.~(\ref{eq:fffinal}).
This procedure is performed  to calculate 
the form-factor products for each $q^2$ bin independently.
The correlation between the $q^2$ bins is negligible due to the excellent $q^2$ resolution.

The  procedure described above provides the form-factor products
with an arbitrary normalization factor common for all of them.
In this work we use the normalization
$q^2 |H_0(q^2)|^2 \to 1$ when $q^2 \to 0$.

In total, 16181 $D^{+} \to K^{-} \pi^+ e^+ \nu_e$ candidates
are selected in the $K^*$-dominated region.
The influence of the small residual background 
on the results is insignificant.
To avoid numerical instability caused by negative bin content 
after background subtraction, the final results presented in Table~\ref{tab:pwt_result} are 
obtained neglecting the background contribution.

\begin{figure}[htp]
  \begin{center}
  
\includegraphics[width=\linewidth]{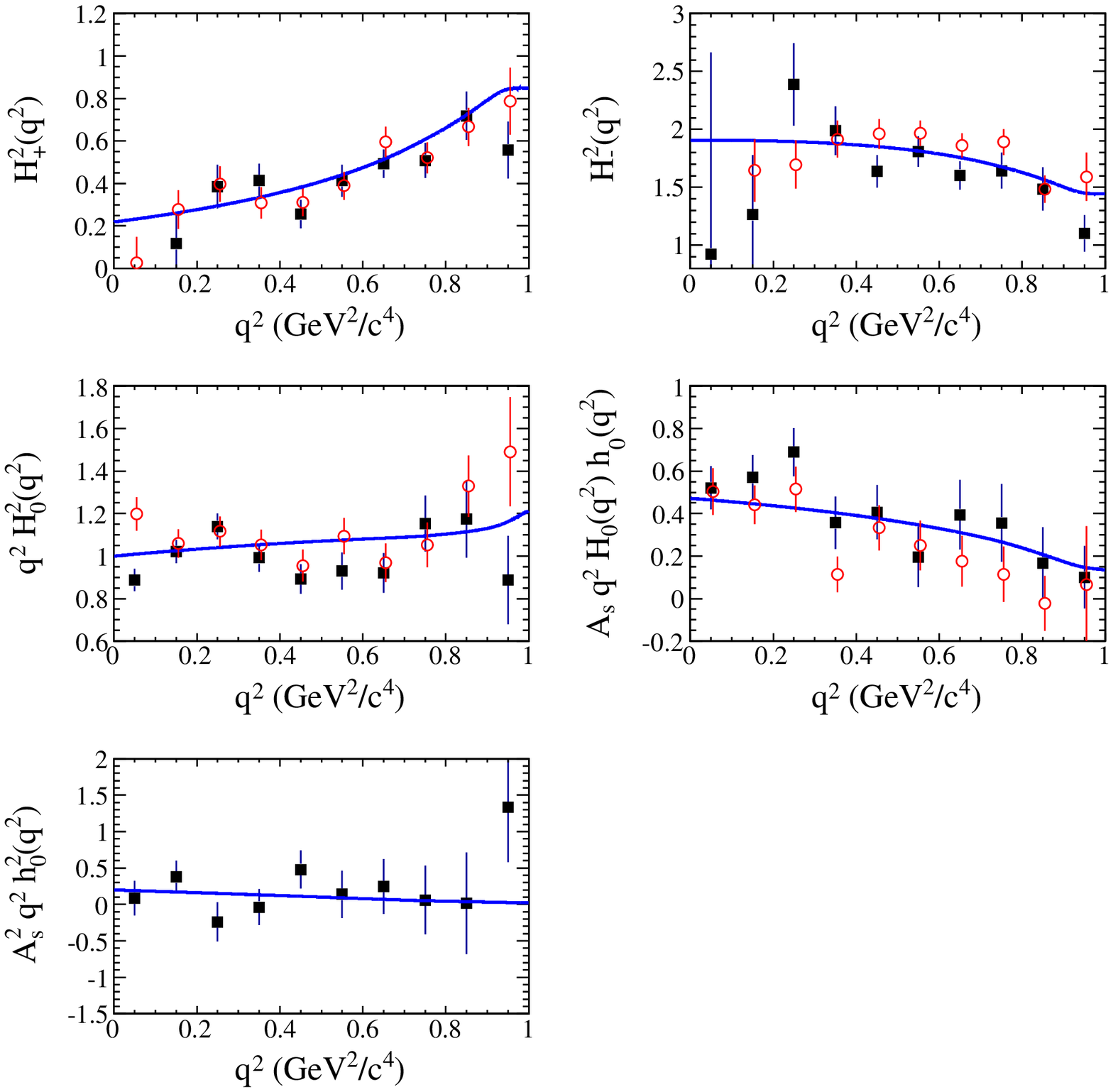}
  \caption{ Average form-factor products in the $K^{*}$-dominated region. 
  The model-independent measurements in this work (squares) are
compared with the CLEO-c results (circles) and with our PWA solution (curves).
In the CLEO-c results, 0.33 GeV$^{-1}$ is taken as the $A_{S}$ value for 
comparison~\cite{Link:2002wg}.
  Error bars represent statistical and systematic uncertainties combined in quadrature.}
  \label{fig:pwt_result_cleo_pwa}
  \end{center}
\end{figure}

In \figref{fig:pwt_result_cleo_pwa} the results are compared with
the CLEO-c results~\cite{Briere:2010zc} and with 
our PWA solution.
The model-independent measurements are consistent with
the SPD model with the parameters determined by the PWA fit.
They are also consistent with the results previously reported
by CLEO-c.

\begin{table*}[htp]
\small
  \begin{center}
  \renewcommand{\arraystretch}{1.15}
  \renewcommand{\tabcolsep}{0.01\linewidth}
  \caption{Average form-factor products in the $K^{*}$-dominated region. The first 
  and second uncertainties are   statistical and systematic, respectively.}
  \begin{tabular}{l|c c c c c}
  \hline
  \hline
$q^2$ (GeV$^2/c^4)$ &  $H^{2}_{+}(q^{2})$ & $H^{2}_{-}(q^{2})$ & $q^{2} H^{2}_{0}(q^{2})$ & $A_{s} q^{2} H_{0}(q^{2}) h_{0}(q^{2})$ & $A_{s}^{2} q^{2} h_{0}^{2}(q^{2})$  \\ \hline  \hline
$0.0~-~0.1$   & $1.67 \pm 0.46 \pm 0.12$   & $0.92 \pm 1.71 \pm 0.31$   & $0.89 \pm 0.05 \pm 0.02$   & $0.52 \pm 0.08 \pm 0.06$   & $0.09 \pm 0.23 \pm 0.05$   \\ 
$0.1~-~0.2$   & $0.12 \pm 0.13 \pm 0.05$   & $1.26 \pm 0.50 \pm 0.12$   & $1.02 \pm 0.05 \pm 0.02$   & $0.57 \pm 0.09 \pm 0.05$   & $0.38 \pm 0.21 \pm 0.05$   \\ 
$0.2~-~0.3$   & $0.39 \pm 0.10 \pm 0.03$   & $2.39 \pm 0.33 \pm 0.13$   & $1.14 \pm 0.06 \pm 0.02$   & $0.69 \pm 0.10 \pm 0.05$   & $-0.24 \pm 0.24 \pm 0.11$   \\ 
$0.3~-~0.4$   & $0.41 \pm 0.07 \pm 0.03$   & $1.99 \pm 0.20 \pm 0.07$   & $0.99 \pm 0.06 \pm 0.03$   & $0.36 \pm 0.10 \pm 0.07$   & $-0.04 \pm 0.23 \pm 0.10$   \\
$0.4~-~0.5$   & $0.26 \pm 0.06 \pm 0.03$   & $1.64 \pm 0.13 \pm 0.06$   & $0.89 \pm 0.06 \pm 0.04$   & $0.41 \pm 0.11 \pm 0.06$   & $0.48 \pm 0.22 \pm 0.14$   \\ 
$0.5~-~0.6$   & $0.41 \pm 0.06 \pm 0.05$   & $1.81 \pm 0.11 \pm 0.07$   & $0.93 \pm 0.07 \pm 0.05$   & $0.20 \pm 0.12 \pm 0.07$   & $0.14 \pm 0.27 \pm 0.18$   \\ 
$0.6~-~0.7$   & $0.49 \pm 0.06 \pm 0.03$   & $1.60 \pm 0.10 \pm 0.07$   & $0.92 \pm 0.08 \pm 0.05$   & $0.39 \pm 0.14 \pm 0.09$   & $0.25 \pm 0.31 \pm 0.22$   \\ 
$0.7~-~0.8$   & $0.51 \pm 0.06 \pm 0.05$   & $1.64 \pm 0.10 \pm 0.12$   & $1.15 \pm 0.10 \pm 0.09$   & $0.36 \pm 0.15 \pm 0.11$   & $0.06 \pm 0.39 \pm 0.27$   \\ 
$0.8~-~0.9$   & $0.72 \pm 0.08 \pm 0.08$   & $1.49 \pm 0.11 \pm 0.15$   & $1.17 \pm 0.11 \pm 0.15$   & $0.17 \pm 0.14 \pm 0.10$   & $0.02 \pm 0.56 \pm 0.42$   \\
$0.9~-~1.0$   &  $0.56 \pm 0.13 \pm 0.01$   &   $1.10 \pm 0.15 \pm 0.05$   &   $0.89 \pm 0.18 \pm 0.11$   &  $0.10 \pm 0.14 \pm 0.03$   &   $1.33 \pm 0.67 \pm 0.33$   \\ \hline
 \end{tabular}
  \label{tab:pwt_result}
  \end{center}
\end{table*}

\begin{table*}[htp]
\small
  \begin{center}
  \renewcommand{\arraystretch}{1.15}
  \caption{Systematic uncertainties of the form-factor products:
  the first numbers are uncertainties due to the limited PHSP sample size, while the second represent uncertainties due to    the $\vec{m}_{\alpha}$ calculation.}
  \begin{tabular}{l|x{1.3cm} x{1.3cm}|x{1.3cm} x{1.3cm}|x{1.3cm} x{1.3cm}|x{1.3cm} x{1.3cm}|x{1.3cm} x{1.3cm}} 
  \hline \hline
$q^2$ (GeV$^2/c^4$) &  
\multicolumn{2}{c|}{$H^{2}_{+}(q^{2})$} &
\multicolumn{2}{c|}{$H^{2}_{-}(q^{2})$} & 
\multicolumn{2}{c|}{$q^{2} H^{2}_{0}(q^{2})$} &
\multicolumn{2}{c|}{$A_{s} q^{2} H_{0}(q^{2}) h_{0}(q^{2})$} &
   \multicolumn{2}{c}{$A_{s}^{2} q^{2} h_{0}^{2}(q^{2})$}  \\ \hline  \hline
$0.0~-~0.1$   & $0.11$   & $0.05$   & $0.14$   & $0.27$   & $0.02$   & $0.00$   & $0.05$   & $0.03$   & $0.04$   & $0.02$   \\ \hline
$0.1~-~0.2$   & $0.05$   & $0.03$   & $0.07$   & $0.10$   & $0.02$   & $0.00$   & $0.05$   & $0.01$   & $0.05$   & $0.01$   \\ \hline
$0.2~-~0.3$   & $0.03$   & $0.01$   & $0.06$   & $0.11$   & $0.02$   & $0.00$   & $0.05$   & $0.00$   & $0.07$   & $0.08$   \\ \hline
$0.3~-~0.4$   & $0.03$   & $0.01$   & $0.06$   & $0.05$   & $0.03$   & $0.02$   & $0.06$   & $0.03$   & $0.09$   & $0.05$   \\ \hline
$0.4~-~0.5$   & $0.03$   & $0.01$   & $0.06$   & $0.02$   & $0.03$   & $0.02$   & $0.06$   & $0.01$   & $0.12$   & $0.06$   \\ \hline
$0.5~-~0.6$   & $0.03$   & $0.03$   & $0.07$   & $0.01$   & $0.05$   & $0.03$   & $0.07$   & $0.02$   & $0.17$   & $0.06$   \\ \hline
$0.6~-~0.7$   & $0.03$   & $0.01$   & $0.06$   & $0.04$   & $0.05$   & $0.03$   & $0.08$   & $0.04$   & $0.17$   & $0.14$   \\ \hline
$0.7~-~0.8$   & $0.04$   & $0.04$   & $0.08$   & $0.08$   & $0.05$   & $0.07$   & $0.08$   & $0.07$   & $0.26$   & $0.02$   \\ \hline
$0.8~-~0.9$   & $0.06$   & $0.05$   & $0.10$   & $0.11$   & $0.08$   & $0.12$   & $0.09$   & $0.03$   & $0.41$   & $0.03$   \\ \hline
$0.9~-~1.0$   & $0.01$   &   $0.01$   &   $0.01$   &   $0.05$   &   $0.01$   &  $0.11$   &  $0.01$   &  $0.03$   &  $0.04$   &  $0.33$   \\ \hline
\end{tabular}
  \label{tab:pwt_syst}
  \end{center}
\end{table*}

The systematic uncertainties of the form-factor product determination
originate mostly from  the $\vec{m}_{\alpha}$  calculation.
They are estimated using a large generator-level PHSP sample, 
with which the form-factor products are computed using the generator-level kinematic variables.
The difference between the input and the computed value is taken as 
the systematic uncertainty related to the $\vec{m}_{\alpha}$ calculation procedure.
The limited statistics of PHSP signal MC used to calculate the
 $\vec{m}_{\alpha}$ vectors is another source of uncertainty. To estimate its contribution,
we randomly select subsamples from the generator-level PHSP sample with roughly the size of the PHSP signal MC. 
The standard deviation of the form-factor products computed using the 
different subsamples is taken as the systematic uncertainty.
%
%
The uncertainties
due to neglecting the residual background as well as  from other sources are negligible.
The main systematic uncertainties are presented in Table~\ref{tab:pwt_syst}.

\section{Summary}

An analysis of $D^+ \to K^- \pi^+ e^{+} \nu_{e}$ has been performed
and its branching fraction
has been measured over the full $m_{K\pi}$ range (0.6$<$$m_{K\pi}$$<$1.6 GeV/$c^{2}$)
and in the $K^*$-dominated 
region  (0.8$<$$m_{K\pi}$$<$1.0 GeV/$c^{2}$).

Using a PWA fit, we have analyzed the components in the $D^+ \to K^- \pi^+ e^{+} \nu_{e}$ decay.
In addition to the process $D^{+} \to \bar{K}^{*}(892)^{0} e^{+} \nu_{e}$, 
we observed the $K\pi$ \emph S-wave component with a fraction of $(6.05\pm0.22\pm0.18)\%$. 
Possible contributions from the $\bar K^{*}(1410)^{0}$ and $\bar K_{2}^{*}(1430)^{0}$ 
were observed to have significances less than $5\,\sigma$
and the upper limits were provided.

With the signal including the \emph S-wave and $\bar K^{*}(892)^{0}$ as the nominal fit, 
the form factors based on the SPD model, 
together with the parameters describing 
the $\bar K^{*}(892)^{0}$, were measured.
We performed the first measurement of
the vector pole mass $m_{V}$ in this decay,
$m_{V}=1.81^{+0.25}_{-0.17}\pm0.02\, \text{GeV}/c^{2}$.
In the channel $D^{0}\to K^{-}e^{+}\nu_{e}$,
the value $m_{V}=1.884\pm0.012\pm0.014 \,\text{GeV}/c^{2}$ was obtained~\cite{mV kenu}.
When we fixed $m_{V}$ at 2.0 GeV/$c^{2}$ as in Ref.~\cite{pwa2},
consistent results for the form factor parameters were obtained,
as shown in Table~\ref{tab:mV_compare}.

We measured the \emph S-wave phase variation with $m_{K\pi}$ in a 
model-independent way, and found an agreement 
with the PWA solution based on the parameterization in the LASS scattering 
experiment. 

Finally, we performed a model-independent measurement of the $q^{2}$ dependence 
of the helicity basis form factors. It agreed well with the CLEO-c result and 
the PWA solution based on the SPD model.

\begin{table}[htp]
  \begin{center}
  \renewcommand{\arraystretch}{1.15}
  \caption{
 Form factor parameter results
 with $m_{V}$ allowed to vary or fixed at 2.0 GeV/$c^{2}$.
   The first and second uncertainties are
statistical and systematic, respectively. When $m_{V}$ is fixed,
 the $m_{V}$ induced   uncertainty is especially considered by 
   varying $m_{V}$ from 1.7 to 2.2 GeV/$c^{2}$
 besides the ones  listed in Table~\ref{tab:pwasyst}.}

  \begin{tabular}{p{0.27\linewidth}    p{0.38\linewidth}   p{0.32\linewidth} }
  \hline
  \hline
  Variable & $m_{V}$ allowed to vary & $m_{V}$ fixed  \\
  \hline 
$m_{V}$ (GeV/$c^{2}$) & 1.81$^{+0.25}_{-0.17}\pm$0.02   &  2.0  \\
$m_{A}$ (GeV/$c^{2}$) & 2.61$^{+0.22}_{-0.17}\pm$0.03 & 2.64$^{+0.22}_{-0.17}\pm$0.07  \\
$r_{V}$ & 1.411$\pm$0.058$\pm$0.007 & 1.449$\pm$0.034$\pm$0.071  \\
$r_{2}$ & 0.788$\pm$0.042$\pm$0.008 & 0.795$\pm$0.040$\pm$0.016  \\
$A_{1}(0)$ & 0.589$\pm$0.010$\pm$0.012  &  0.589$\pm$0.010$\pm$0.014 \\
  \hline  
  \hline
  \end{tabular}
  \label{tab:mV_compare}
  \end{center}
\end{table}

\section{ACKNOWLEDGEMENTS}
The BESIII collaboration thanks the staff of BEPCII and the IHEP computing center for their strong support.
This work is supported in part by National Key Basic Research Program of China under Contract Nos. 2009CB825204, 2015CB856700;
National Natural Science Foundation of China (NSFC) under Contracts Nos. 10935007, 11075174, 11121092, 11235011,  11125525, 11322544, 11335008, 11425524, 11475185;
the Chinese Academy of Sciences (CAS) Large-Scale Scientific Facility Program;
the CAS Center for Excellence in Particle Physics (CCEPP);
the Collaborative Innovation Center for Particles and Interactions (CICPI);
Joint Large-Scale Scientific Facility Funds of the NSFC and CAS under Contracts Nos. 11179007, U1232201, U1332201;
CAS under Contracts Nos. KJCX2-YW-N29, KJCX2-YW-N45;
100 Talents Program of CAS;
National 1000 Talents Program of China;
INPAC and Shanghai Key Laboratory for Particle Physics and Cosmology;
German Research Foundation DFG under Contract No. Collaborative Research Center CRC-1044;
Istituto Nazionale di Fisica Nucleare, Italy;
Ministry of Development of Turkey under Contract No. DPT2006K-120470;
Russian Foundation for Basic Research under Contract No. 14-07-91152;
The Swedish Resarch Council;
U. S. Department of Energy under Contracts Nos. DE-FG02-04ER41291, DE-FG02-05ER41374, DE-SC0012069, DESC0010118;
U.S. National Science Foundation;
University of Groningen (RuG) and the Helmholtzzentrum fuer Schwerionenforschung GmbH (GSI), Darmstadt;
WCU Program of National Research Foundation of Korea under Contract No. R32-2008-000-10155-0.

\end{document}